\documentclass[12pt]{article}
\usepackage[left=2cm, right=2cm, bottom=2cm ,top = 2cm, footskip=1cm]{geometry}
\usepackage{sectsty}
\usepackage{graphicx}
\usepackage[dvipsnames]{xcolor}
\usepackage[T1]{fontenc}
\usepackage[utf8]{inputenc}
\usepackage{mathtools}
\usepackage{amsmath}
\usepackage{hyperref}
\usepackage{graphicx, amsmath, amsthm, amsfonts, bigints}
\usepackage[labelfont=bf,textfont=md]{caption}

\usepackage{float}

\usepackage{lineno}
\usepackage{chngcntr}

\usepackage{placeins}

\newcommand{\R}{\mathbb{R}}

\begin{document}

\title{Fluctuations in EEG band power at subject-specific timescales over minutes to days explain changes in seizure evolutions}
\author{Mariella Panagiotopoulou$^{1}$, Christoforos A Papasavvas$^{1}$, Gabrielle M Schroeder$^{1}$,\\ Rhys H Thomas$^{2}$, Peter N Taylor$^{1,2,3}$, Yujiang Wang$^{*1,2,3}$}
\date{\today}

\maketitle 

\begin{enumerate}
\item{CNNP Lab (www.cnnp-lab.com), Interdisciplinary Computing and Complex BioSystems Group, School of Computing, Newcastle University, Newcastle upon Tyne, United Kingdom}
\item{Faculty of Medical Sciences, Newcastle University, Newcastle upon Tyne, United Kingdom}
\item{UCL Queen Square Institute of Neurology, Queen Square, London, United Kingdom}
\end{enumerate}
\begin{center}
* Yujiang.Wang@newcastle.ac.uk    
\end{center}




\begin{abstract}

Epilepsy is recognised as a dynamic disease, where both seizure susceptibility and seizure characteristics themselves change over time. Specifically, we recently quantified the variable electrographic spatio-temporal seizure evolutions that exist within individual patients. This variability appears to follow subject-specific circadian, or longer, timescale modulations. It is therefore important to know whether continuously-recorded interictal iEEG features can capture signatures of these modulations over different timescales. 

In this work, we analyse continuous intracranial electroencephalographic (iEEG) recordings from video-telemetry units and find fluctuations in iEEG band power over timescales ranging from minutes up to twelve days.

As expected and in agreement with previous studies, we find that all subjects show a circadian fluctuation in their iEEG band power. We additionally find other fluctuations of similar magnitude on subject-specific timescales. Importantly, we find that a combination of these fluctuations on different timescales can explain changes in seizure evolutions in most subjects above chance level.

These results suggest that subject-specific fluctuations in iEEG band power over timescales of minutes to days may serve as markers of seizure modulating processes. We hope that future work can link these detected fluctuations to their biological driver(s). There is a critical need to better understand seizure modulating processes, as this will enable the development of novel treatment strategies that could minimise the seizure spread, duration, or severity and therefore the clinical impact of seizures.

\end{abstract}

\newpage
\section{Introduction}

Epilepsy is a common neurological condition characterised by recurrent, unprovoked seizures \cite{fisher_ilae_2014}. It affects approximately 1\% of the world's population and a third of patients experience refractory epilepsy, where seizures are not adequately controlled despite medication\cite{chen_treatment_2018}. 

Importantly, epilepsy is not a static disorder; electrographic seizure and epileptiform activities have been shown to fluctuate over hours to years in both intensity and spatial distribution. Specifically, while seizures often share common features in the same patient \cite{kramer_coalescence_2010, schindler_forbidden_2011,schevon_evidence_2012,burns_network_2014, wagner_microscale_2015,truccolo_single-neuron_2011}, electrographic seizure activity may change in terms of duration \cite{cook_human_2016}, spatial spread \cite{marciani_effects_1986,karthick_prediction_2018,naftulin_ictal_2018, pensel_predictors_2020}, spectral properties \cite{alarcon_power_1995} from one seizure to the next. Our recent work \cite{schroeder_seizure_2020} has additionally shown that the seizure EEG spatio-temporal evolution from seizure start to seizure termination (or short: ``seizure evolution'') also changes from one seizure to the next in the same patient. Notably, these changes were consistent with daily (circadian) and/or longer-term fluctuations in most patients \cite{schroeder_seizure_2020}. In support of our observations, a recent study quantifying single-channel properties of seizure onset and offset also noted that different types of dynamics can be seen across different seizures in the same patient \cite{saggio_taxonomy_2020}. Similarly, seizure symptoms are also known to change over time. For example, focal seizures, which evolve into bilateral tonic-clonic seizures, preferentially arise from sleep \cite{jobst_secondarily_2001}. Subclinical seizures (without clinical symptoms) are also reported to follow circadian patterns \cite{jin_prevalence_2017}. Finally, seizure severity appears to depend on the severity of the preceding seizure in the same patient \cite{sunderam_epileptic_2007}. Thus, epileptic seizures are not a fully deterministic sequence of abnormal brain activity patterns, but are clearly modulated by processes that shape the neural activity during a seizure and affect seizure severity.

However, it is unclear what these seizure-modulating processes are, and how to quantify and measure them. Given the evidence of seizure properties fluctuating over various timescales of hours to days, we hypothesise here that the seizure-modulating processes will also fluctuate over these timescales. From existing literature, we also know that continuously recorded electroencephalograms (EEG) show fluctuations over such timescales. For example, spectral properties of the EEG change from moment to moment \cite{oken_short-term_1988} and also follow a circadian rhythm \cite{aeschbach_two_1999}. Global and local characteristics of the continuously recorded (interictal) functional network fluctuate over timescales from hours to days, with circadian rhythm having a particularly strong effect on these dynamics \cite{geier_time-dependent_2015, geier_long-term_2017, mitsis_functional_2020}. Interictal fluctuations related to epilepsy are also seen: high frequency oscillation (HFO) rates vary in location and power within each subject over time \cite{gliske_variability_2018}. Interictal spikes also change in their location and rate over hours to days \cite{karoly_interictal_2016,gliske_variability_2018,conrad_spatial_2020,baud_multi-day_2018,chen_spatiotemporal_2020}. 

We therefore hypothesised that fluctuations of certain features captured in continuously recorded EEG may serve as biomarkers of seizure-modulating processes. We expected these fluctuations to appear on the timescale of hours to days, and we investigated if they can also explain how seizure evolutions change within the same patient. Previous work suggests that many interictal features, including interictal spike rate \cite{baud_multi-day_2018, proix_forecasting_2021, karoly_interictal_2016,karoly_circadian_2017} and high frequency oscillation rate \cite{gliske_variability_2018,scott_viability_2021, chen_spatiotemporal_2020} may serve as biomarkers for modulatory processes. Here, we investigate the full spectral range, using band power in main EEG frequency bands, to capture a complete view of brain activities. Specifically, we use clustering and dimensionality reduction to detect subject-specific spectral patterns in continuously recorded EEG. We then extract the temporal fluctuations over minutes, hour, and days in these common spectral patterns and explore whether fluctuations on different timescales are associated with how seizure evolutions change in each subject.

\section{Methods\label{sec:methods}}

\subsection{Data acquisition\label{subsec:data_acquisition}}

We analysed open source data from subjects with drug-resistant focal epilepsy (available at \href{http://ieeg-swez.ethz.ch}{http://ieeg-swez.ethz.ch}) in accordance with the ethical standards set by the Newcastle University Ethics Committee (Ref: 18818/2019). The data consist of a total of 2656 hours of long-term intracranial electroencephalography (iEEG) from 18 subjects. Continuous recordings in each subject cover 24 to 128 EEG channels and vary between 2 to 12 days. More information about the data is given in \ref{suppl:tableData}. Sampling frequency was either 512 or 1024 Hz depending on the subject. Electrodes (strip, grid, and depth) were implanted intracranially by clinicians. The collection of the data was conducted in the Sleep-Wake-Epilepsy-Center (SWEC) at the University Hospital of Bern, Department of Neurology, as part of their presurgical evaluation programme, independently of this study \cite{burrello_laelaps:_2019}.\\

The iEEG signals were provided in already preprocessed form. Briefly, signals were median-referenced and band-pass filtered from 0.5-120 Hz using a $4^{th}$ order Butterworth filter (forward and backward). Seizure onset and termination times were determined by a board-certified epileptologist. Channels with artifacts were also identified and excluded by the same epileptologist. These steps were all conducted independently of this study and resulted in the publicly available data and annotations. All subjects formally consented to their iEEG data being used for research purposes. \cite{burrello_laelaps:_2019}.

\begin{figure}[ht!]
    \hspace{-0.8cm}
    \includegraphics[scale = 1]{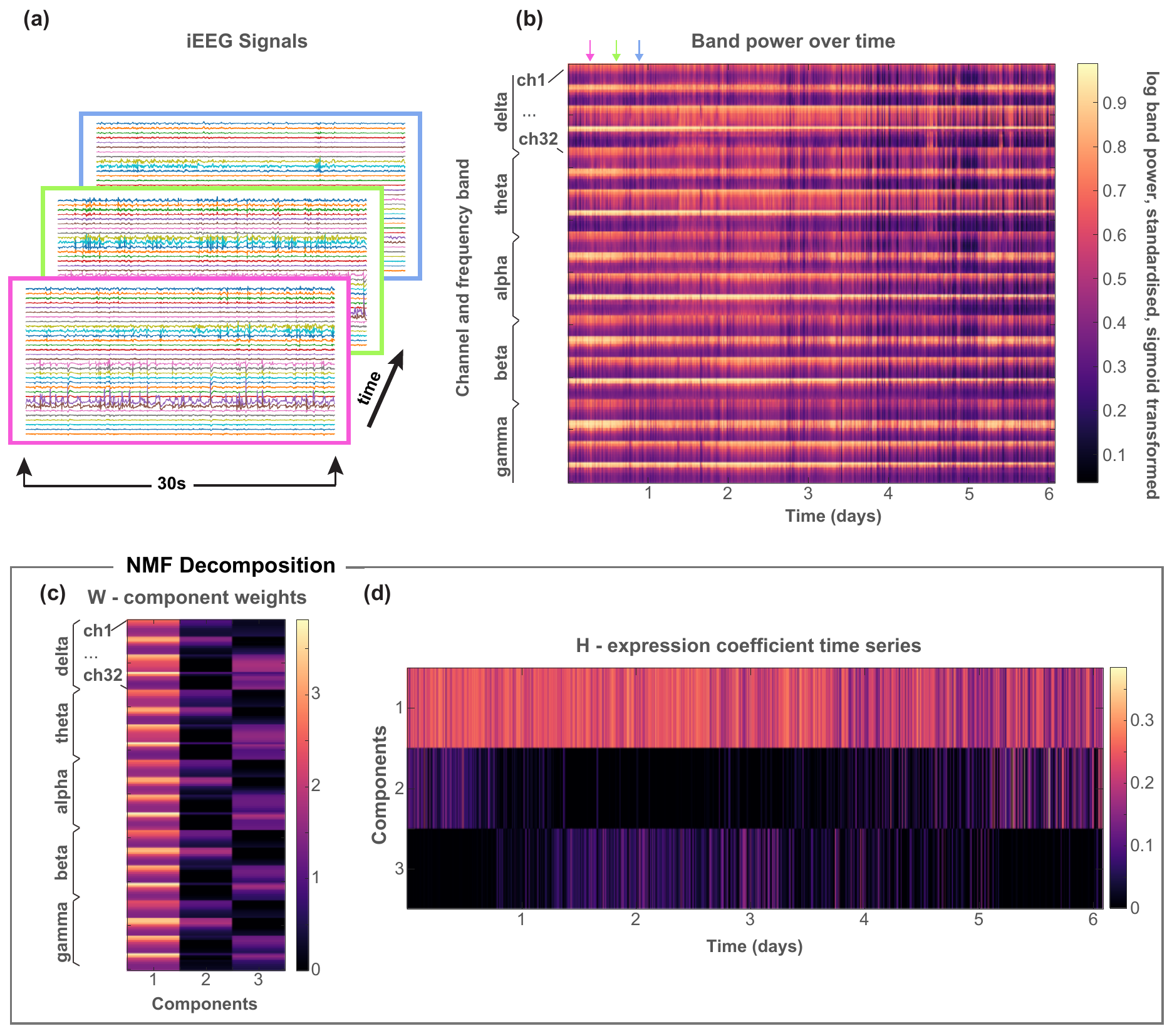}
      \caption{\textbf{Workflow of data preprocessing; calculation of band power in 30~s epochs and subsequent dimensionality reduction to detect subject-specific spectral patterns}. (a) The multi-channel continuous iEEG recording was divided into 30~s non-overlapping epochs. (b) The standardised, log and sigmoid transformed band power. (c)\&(d) NMF (dimensionality reduction) of the band power matrix results in the decomposition $W \times H$. (c) The matrix $W$ contains the basis vectors, each of which had $5 \times \#channels$ weights that represents a pattern of frequency across all channels and frequency bands. (d) The coefficients matrix $H$ captures the contribution of each frequency pattern (basis vector) to each time window.}
    \label{fig:NMF_illustration}
\end{figure}

\subsection{IEEG preprocessing\label{subsec:preprocessing}}

We performed additional preprocessing steps to extract iEEG band power from five main frequency bands (Fig.~\ref{fig:NMF_illustration}a). For each recording channel, the signal was divided into 30~s epochs (Fig.~\ref{fig:NMF_illustration}b). For each epoch, the band power was computed for the following frequency bands: $\delta:~1-4$ Hz, $\theta:~4-8$ Hz, $\alpha:~8-13$ Hz, $\beta:~13-30$ Hz and $\gamma:~30-80$ Hz. Band power across the five main frequency bands was estimated using Welch's method for every 30~s epoch, with 3~s sliding window without overlap between consecutive windows. This estimation yielded a time-varying band power, with each time point corresponding to the mean power within a 30~s window. The band power values were aggregated into band-specific matrices with dimensions \#channels $\times$ \#epochs. Then, these matrices were log transformed and standardised across all epochs and channels within a frequency band. To enable subsequent analysis steps, we also Sigmoid transformed ($S(x)=[1+\exp(-x)]^{-1}$) the standardised data to ensure positive entries between 0 and 1. For each subject, we then concatenated the matrices from all frequency bands yielding a single (5 $\times$ \#channels) $\times$ \#epochs (henceforth defined as $n \times T$; Fig.~\ref{fig:NMF_illustration}b). We will refer to this matrix as the data matrix $X$ throughout the paper.

Note that we did not exclude seizure epochs from the construction of the data matrix, as seizures only represent a few epochs in the context of the continuous recording. Our downstream analysis (with empirical mode decomposition) is robust to ``noise'' of this type, and we show in \ref{subsec:Raw_analysis-noise} that this choice does not affect our main results. Note also that our measure of how seizure evolutions change over time (in Section \ref{subsec:sz_diss}) \cite{schroeder_seizure_2020} was based on functional network activity of the seizure, whilst we used band power features to measure fluctuations of the continuous EEG. Therefore, the fluctuations of the continuous EEG were not trivially related to variability in seizure evolutions (also shown in \ref{suppl:Corr_band_power}).

    

\subsection{Non-Negative Matrix Factorization for dimensionality reduction}\label{subsec:NMF}

As the data matrix $X$ for each subject is high-dimensional with redundant information (e.g. in different channels), we applied a dimensionality reduction step on $X$ using Non-Negative Matrix Factorization (NMF) \cite{lee_learning_1999}. NMF provides a low-rank approximation to a non-negative input matrix $X\in \R_{+}^{n\times T}$ as the product of two non-negative matrices, $W\in \R_{+}^{n \times k}$ and $H\in \R_{+}^{k\times T}$, such that $X \approx W \times H \equiv X^{'}$, given an integer $k$. Specifically, we applied the non-negative singular value decomposition (SVD) with low-rank correction (NNSVD-LRC) \cite{atif_improved_2019}, which is a method of low-rank approximation using an NMF initialisation approach based on SVD.  

In this way, we decomposed each subject's band power data matrix $X$ into $W$ and $H$ matrices (Fig.~\ref{fig:NMF_illustration}c-d). Every column of matrix $W$ corresponded to a single NMF component and formed a basis vector or feature weight with $n$ elements. Each row of $H$ represented how a single NMF component evolves over time across all $T$ time epochs. We also refer to a single row of $H$ as the NMF-expression coefficient time series. This dimensionality reduction step not only compressed the data matrix $X$ into few relevant dimensions, but can also be understood as a data-driven pattern detection, or (soft) clustering of recurrent spectral patterns in the continuously recorded EEG. For example, Fig.~\ref{fig:NMF_illustration} shows that the the band power in each channel at a particular time window could be (approximately) described as a weighted sum of three patterns (given by the three basis vectors in $W$). The weights were given as the expression coefficients (in $H$) at each time point.
This data-driven spectral pattern detection essentially provided us with a comprehensive view of the EEG in each subject, without the need to pre-define specific patterns of interest, which may not acknowledge subject-specific variations in these spectral patterns.


To determine the optimal number of representative NMF components, $k$, for each subject, we performed NNSVD-LRC for $k = 3, 4, \dots, 15$. For each value of $k$, we obtained the matrices $W$ and $H$. Using these matrices, we calculated the relative reconstruction errors $\sum_{n,T}|X-X'|/(n \times T)$, as well as the quantity $c=max\{max(|Corr(W)|), max(|Corr(H)|)\}$ for each $k$, where $max(|Corr(W)|)$ represents the maximum absolute correlation among all column pairs of $W$, and $max(|Corr(H)|$ represents the maximum absolute correlation among all row pairs of $H$. The latter represents the strongest correlation or anticorrelation between NMF components in terms of their feature weights $W$ and their expression coefficient time series $H$. In this way, redundant information, particularly in $H$, was excluded whilst preserving the important spatio-temporal patterns for the next processing steps. 
A distinct number of NMF components, $k$, was selected for each subject. This was the $k$ yielding the smallest correlation between the NMF components that had a relative reconstruction error smaller than $5\%$.
    
After determining the optimal choice of $k$, we obtained two matrices for each subject, $W$ and $H$. To re-iterate, the matrix $W$ consists of the basis vectors, while $H$ is a multivariate time series with dimensions equal to $k \times T$ ($=$ the number of NMF components $\times$ the total number of epochs).

\subsection{Extracting fluctuations in interictal band power using MEMD\label{subsec:MEMD}}

To investigate fluctuations in band power on different timescales, we analysed the matrix $H$ using Empirical Mode Decomposition (EMD) \cite{huang_empirical_1998, huang_confidence_2003}. 
It is well known that EEG signals are non-stationary processes characterised by time-varying features \cite{kaplan_nonstationary_2005, fingelkurts_operational_2001}.
EMD is a popular data-adaptive method to detect non-stationary and non-rhythmic fluctuations on different timescales. Compared to Fourier and Wavelet-based approaches, EMD does not assume any particular basis function or local stationarity. EMD also does not require detrended time series and does not make assumptions about trends or timescales of trends. It has the advantage of fully decomposing the signal into the full range of timescales of fluctuations; their point-wise summation fully reconstructs the original signal. As the nature of these band power fluctuations is unknown and most likely not stationary, we opted for a data-driven method that makes as few assumptions as possible.

EMD decomposes an input signal $Y(t)$, into $M$ finite narrow-band fluctuations, known as intrinsic mode functions (IMFs), based on the local extrema of the signal: $Y(t)=\sum_{i=1}^M \text{IMF}_i(t) + r(t)$, where $r(t)$ is the residue signal \cite{huang_empirical_1998}. The IMFs additionally satisfy the properties that make the Hilbert-transform well-defined and therefore naturally yield instantaneous frequency and phases for each IMF.

However, local extrema are not directly applicable to multivariate time series signals \cite{rehman_multivariate_2010}, as we have in the $H$ matrix. Therefore, we used an extension of the EMD to multi-dimensional space, called the Multivariate Empirical Mode Decomposition (MEMD) \cite{rehman_multivariate_2010}. 
In MEMD, multiple projections of the multivariate signal are generated along different directions in n-dimensional spaces; the multidimensional envelope of the signal is then obtained by interpolating across the different envelopes of these projections \cite{rehman_multivariate_2010}.
An additional advantage of this method is that it yields the same number of IMFs across the different dimensions of the multivariate signal, and preserves fluctuations of similar frequency across the different dimensions within each of the IMFs (mode-alignment) \cite{rehman_multivariate_2010}.

For the purpose of this analysis, we used MEMD to decompose the NMF-expression coefficient time series, $H$, into a number of multi-dimensional oscillatory modes. 
Therefore, the matrix $H$ can be represented by the sum of $M$ multi-dimensional IMF signals, where the dimension for each IMF is equal to $k$ (i.e. the number of rows of the matrix $H$, which corresponds to the number of NMF components). To clarify, we can think of all IMFs in a specific dimension as the decomposition of the corresponding NMF-expression coefficient time series. Thus, $\text{IMF}_{i,j}$ refers to the $j$-th dimension of the $i$-th IMF timescale. The $j$-th NMF-expression coefficient time series $H_j=Y_{j}(t)$ can be written as $Y_{j}(t) = \sum_{i=1}^M \text{IMF}_{i,j}(t) + r_{j}(t)$. This equation applies to every NMF component $j = 1, \dots, k$. 

\subsection{Extracting time-varying characteristics from band power fluctuations (IMFs) using Hilbert Spectral Analysis}

To obtain a time-frequency representation of the oscillatory modes (IMFs), and hence derive their time-varying characteristics (instantaneous frequency, phase, and amplitude), we applied a Hilbert-transform on each dimension of the IMF (following classical analysis methods for EMD) \cite{huang_empirical_1998, huang_confidence_2003, huang_hilbert-huang_2014}.


For any (real-valued) univariate signal $u(t)$, we can define its Hilbert transform as:
\begin{equation}
    H(u)(t) = \frac{1}{\pi}P\int_{-\infty}^{+\infty} \frac{u(\tau)}{t-\tau}d\tau,
\end{equation}
where $P$ represents the Cauchy principal value for any function $u(t) \in L^{P}$ class \cite{huang_empirical_1998}. 

The analytical signal $v(t)$ obtained from the Hilbert transform can be expressed as:
\begin{equation}
    v(t) = u(t) + iH(u)(t) = a(t)e^{i\theta(t)},
\end{equation}

where 
\begin{equation}
a(t) = \sqrt{u(t)^2 + H(u)(t)^2}
\end{equation}

and 
\begin{equation}
\theta(t) = \tan^{-1}\left( \frac{H(u)(t)}{u(t)} \right)
\end{equation}

where $a(t)$ and $\theta(t)$ are the instantaneous amplitude and instantaneous phase, respectively.

The instantaneous frequency, $f(t)$, can then be calculated as follows:
\begin{equation}
    f(t) = \frac{d\theta(t)}{dt}.
\end{equation}

The application of EMD along with Hilbert transform leads to the so-called Hilbert-Huang transform. Through the Hilbert spectral analysis, each IMF's instantaneous frequency can be represented as functions of time. The result is a frequency-time distribution of signal amplitude (or energy using the squared values of amplitude, $a^{2}(t)$), designated as Hilbert amplitude spectrum or Hilbert spectrum (or Hilbert energy spectrum if energy is used instead of amplitude), $H(f, t)$.

For each univariate IMF signal, we can obtain the Hilbert energy spectrum as a function of instantaneous frequency and time mathematically using the following formula:

\begin{equation}
    H(f,t)= \left\{
        \begin{array}{ll}
        a^{2}(t), &  f = f(t)\\
        0, & \text{otherwise.}\\
    \end{array} \right. 
    \label{eq:Hi(f,t)}
\end{equation}
For visualisation purposes we will display the inverse of the instantaneous frequency, i.e. the instantaneous period length, also termed `cycle length' in the following. 

The Hilbert-Huang marginal spectrum $h(f)$ of the original signal $u(t)$ can then be defined as the total energy  
distributed across the frequency space within a time period $[0,T]$. Mathematically, this definition can be expressed as shown below:
\begin{equation}
    h(f) = \int_{0}^{T}H(f,t)dt.
    \label{eq:h(f)}
\end{equation}

By using Equations \ref{eq:Hi(f,t)} \& \ref{eq:h(f)} we can obtain the Hilbert-Huang marginal spectrum for a univariate IMF signal. However, the application of the multivariate EMD results in multivariate IMF signals. In order to compute the Hilbert-Huang marginal spectrum of each multivariate IMF signal across all dimensions, we simply averaged over the dimensions $H_{i}(f,t)$ across $i = 1, \dots, k$ dimensions: 

\begin{equation}
    \Bar{H}(f,t) = \frac{\sum_{i=1}^k H_i(f,t)}{k}.
    \label{eq:h1(f)}
\end{equation}

The corresponding marginal spectrum $\Bar{h}(f)$ was then similarly defined as:

\begin{equation}
    \Bar{h}(f) = \int_{0}^{T}\Bar{H}(f,t)dt.
    \label{eq:h2(f)}
\end{equation}

For numerical computations, we discretised time $t$ to compute the integrals as sums. Figure~\ref{fig:MEMD_method} shows the marginal Hilbert-Huang spectra for different multivariate IMFs in an example subject.

\subsection{Peak fluctuation frequency in each IMF\label{subsec:domfreq}}

Within each subject, each IMF was characterised by a peak fluctuation frequency (measured in cycles/day here). It was determined as the frequency with the highest power based on the marginal Hilbert-Huang spectrum over all frequencies, $\Bar{h}(f)$. 

\subsubsection{Finding a circadian IMF \label{subsec:method24hIMF}}
We will later focus one part of our analysis on IMFs that fluctuate on the timescale of 24 hours (1 cycle/day). To detect those IMFs, we found IMF(s) with a peak fluctuation frequency of 1 cycle/day. If two IMFs were found (i.e. both displayed the a peak frequency at around 1 cycle/day), then the circadian IMF was determined to be the IMF with the higher power. This case only occurred in 1 subject. 

\subsection{Relative contribution of iEEG main frequency bands in different IMFs\label{subsec:relative_power}}

To understand how much each of the iEEG frequency bands and channels contributed to a certain IMF, we first determined how much each dimension of the IMF contributed to the overall power of the IMF. To this end, we first obtained the mean power $E_{ij}$ in each dimension $j$ of every $i$-th IMF signal: 
\begin{equation}
    E_{ij} = \frac{\sum_{t=0}^{T} a_{ij}(t)^{2}}{T},
\end{equation}
where $T$ is, as before, the number of time epochs, and $a_{ij}(t)$ is the instantaneous amplitude for the $j$-th dimension of $i$-th IMF signal at time point $t$. One of the main properties of MEMD is that multivariate signals are decomposed into multivariate IMF signals of the same dimensions, where all dimensions within an IMF share fluctuations of the same timescale \cite{lv_multivariate_2016}. Hence, focusing on the mean power over time of each dimension within an IMF is a good indication of the power on a particular timescale. The relative contribution of each $j$-th dimension to the $i$-th IMF (or relative power) was then defined as: 
\begin{equation}\label{eqn:relpower}
    R_{ij} = \frac{E_{ij}}{\sum_{j=1}^{k} E_{ij}},
\end{equation} 
with $k$ indicating the number of dimensions. 

Using the relative contribution of each dimension as weights, we can then form the weighted sum of all dimensions in terms of contributions of iEEG main frequency bands. By summing channel contributions for each iEEG main frequency band (see Fig.~\ref{fig:W_connection}b\&c for an example subject), we obtained a matrix of dimensions (\# main frequency bands $=5$) $\times$ (\# dimensions $=k$). This matrix was then multiplied with the weight indicating the contribution of each dimension to yield a vector (of length \# main frequency bands $=5$) representing the contribution of each main frequency band to particular IMF for each subject (Fig.~\ref{fig:W_connection}d).

\subsection{Different band power fluctuations reveal spatial heterogeneity within iEEG main frequency band \label{subsec:spatial_heterogeneity}}

To determine if all recording channels contribute homogeneously to an IMF in a particular frequency band, we used a measure that quantifies sparsity of a distribution: the \textit{Gini index} \cite{hurley_comparing_2009}. Given a vector $\boldsymbol{x} = (x_{1}, x_{2}, \dots, x_{N})$ sorted in ascending order such that $x_{1} < x_{2} < \dots < x_{N}$, the Gini index can be derived using the following formula:\\
\begin{equation}
    G(\boldsymbol{x}) = 1 - 2\sum_{i=1}^{N} \frac{x_{i}}{\left\|\boldsymbol{x}\right\|_{1}}\left(\frac{N-i+\frac{1}{2}}{N} \right).
\end{equation}

It can range from 0 to 1, with values closer to 0 indicating low sparsity (homogeneity) and values closer to 1 corresponding to higher sparsity (heterogeneity).

We derived the Gini index for each IMF across different channels within each main frequency band. In other words, for each IMF, we first computed the contribution $C_{i}$ to each $i$-th IMF as the product of the relative power (eqn.~\ref{eqn:relpower}) and the weights matrix: $C_{i}=\sum_{j} R_{ij} \times W_{j}$, with $i$ indexing the IMF number, and $j$ indexing its dimension. Specifically, $W_{j}$ is the $j$-th column of the $W$ matrix from the NMF decomposition (Fig.~\ref{fig:NMF_illustration}(c)), whereas $R_{ij}$ is a scalar representing the relative power of the $j$-th dimension to the $i$-th IMF (see \ref{subsec:relative_power}). The resulting $C_i$ is a vector of length \#frequency bands ($=5$) $\times$ \#channels, i.e. the same length as $W_j$. As we are interested in the distribution of each $C_{i}$ across channels for each frequency band, we applied the Gini index to each frequency band separately in each $C_{i}$, yielding one Gini index per frequency band and IMF.

\subsection{Seizure distance in terms of a particular band power fluctuation (IMF)\label{subsec:imf_seizure_distance}}

For each subject, we quantified the difference between pairs of seizures in terms of each IMF. This measure (which we subsequently term the ``IMF distance'') thus quantifies how different two seizures are to each other in terms of a particular fluctuation of the band power. To obtain this difference, we first computed the product $W \times \text{IMF}_{i}(t)$, where $\text{IMF}_{i}(t)$ is the multi-dimensional $i$-th IMF ($k \times T$ matrix). The product yields the matrices $X'_{\text{IMF}_{i}}$ for all $i = 1,\dots, M$ timescales. $X'_{\text{IMF}_{i}}$ reconstructs the $i$-th IMF in the original space of all channels and frequency bands. For each $X'_{\text{IMF}_{i}}$, we computed a distance matrix based on the multivariate Euclidean distance of IMF values for each pair of seizures: $D_i(a,b)=||X'_{\text{IMF}_i}(t_a)-X'_{\text{IMF}_i}(t_b)||$, where $t_a$ and $t_b$ are the time epochs of the seizure pair's onset.
Therefore, we obtained $M$ IMF seizure distance matrices per subject, each representing the pairwise seizure distance for a specific IMF.

Note that any seizure-induced changes in the band power will only be present in a few epochs (as we use 30~s long epochs). Therefore, the seizures are considered to only influence the fastest IMFs (highest-frequency fluctuations), while they have little effect on the slower IMFs. \ref{subsec:raw_analysis_onset-1} additionally shows that our main results were reproduced by using the IMF seizure distances obtained from one epoch before the seizure onset epoch ($t_a-1$ and $t_b-1$).

\subsection{Quantifying differences in seizure evolutions using seizure dissimilarity\label{subsec:sz_diss}}

To quantify how seizures themselves change over time in terms of the seizure EEG evolutions, in our previous work, we introduced a quantitative measure of how dissimilar two seizures are within a subject \cite{schroeder_seizure_2020}. Briefly, each epileptic seizure in a subject was analysed in terms of its evolution through the space of functional network dynamics (using exactly the same pipeline as \cite{schroeder_seizure_2020}). Each pair of seizures was then compared to each other using dynamic time warping \cite{sakoe_dynamic_1978}, allowing us to recognise seizures with shared evolutions (or parts of evolutions), even if the seizures evolved and different rates. The average distance between the warped seizures was then taken as the dissimilarity measure. As such, for each subject, we obtained a ``seizure dissimilarity'' matrix, which captures the pairwise dissimilarity between the subject's seizure evolutions.

\subsection{Association between seizure dissimilarity \& IMF seizure distance}\label{subsec:Linear analysis}

Finally, we related how seizure evolutions changed over time (quantified using seizure dissimilarity) with fluctuations seen in the continuously recorded iEEG (quantified using IMF seizure distances). In subjects with at least six recorded seizures, we investigated if IMF seizure distances were associated with seizure dissimilarity. For every subject, we used a linear regression framework, where the seizure dissimilarity was the response variable and the IMF seizure distances were the explanatory variables. The observations were the entries of the seizure dissimilarity matrix and IMF seizure distance matrix. As each matrix was symmetric, we only used the upper/lower triangular elements. We also included the EMD residue signal distances, and temporal distances of seizures (how far apart in time each pair of seizures occurred) as additional explanatory variables. 
The response, as well as the explanatory variables, were standardised individually before fitting the model.

We performed a variable selection step for our analysis, as the number of explanatory variables (i.e. $M+2$) was relatively large. We used LASSO (Least Absolute Shrinkage and Selection Operator)\cite{tibshirani_regression_1996}, which is a sparse shrinkage method. Linear regression coefficients were calculated based on least squares, subject to the $L_{1}$ penalty. The LASSO also accounted for any collinearity issues between variables. As we were interested in detecting positive relationships between the response variable (as these were distances) and the explanatory variables, we used a constrained positive LASSO; that is, coefficients were constrained to be non-negative. For the LASSO, the tuning parameter $\lambda$ was selected using a 10-fold cross validation method from a range of values $\lambda = 10^{-3}, 10^{-2.95}\dots, 10^{1.95}, 10^{2}$ (see Fig.~\ref{fig:cv_error}). 

After selecting a small number of explanatory variables, an ordinary least squares regression was performed for each subject to obtain $R^2$ and $95\%$ confidence intervals for the coefficients.

\subsection{Statistical analysis\label{subsec:statistical_analysis}}

To assess if the level of explanatory power of the best model selected for each subject has occurred by chance, we performed two separate tests of statistical significance for the adjusted $R^2$. Both test yielded very similar results and are are shown in \ref{suppl:signtest}.

In the first test, we randomly selected seizure onset times by generating a sample from the uniform distribution on the interval $(0,T)$ over 500 iterations. The size of the sample was equal to the number of annotated seizures for each iteration.  
Then, keeping the randomly picked seizure onset times unsorted, we obtained for each one of them new IMF seizure distance matrices and performed the LASSO and linear regression, as described in the previous section, leaving the response variable unchanged. Finally, we calculated the adjusted $R^2$ for each iteration. Across all iterations, the adjusted $R^2$ values were used to estimate the distribution of the test statistic used in the permutation test. P-values were then calculated as the percentage of adjusted $R^2$ values that were larger in the permutation distribution. Statistical significance was determined based on a significance level of $5\%$.  

In the second test, we permuted the order of the seizures without permuting the seizure timing over 500 iterations. We then performed the LASSO and subsequent steps as in the first test.

\subsection{Data and code availability\label{subsec:data_code_availability}}

The long-term iEEG recordings for all subjects are
available at \href{http://ieeg-swez.ethz.ch/}{http://ieeg-swez.ethz.ch/} under the section ``Long-term Dataset and Algorithms'' \cite{burrello_laelaps:_2019}. 

Initial signal processing was performed using Matlab version 2019a and Matlab's built-in functions. 
NMF and MEMD were implemented using the following publicly available functions:
\begin{itemize}
    \item \textbf{Non-negative matrix factorisation} was conducted using the \texttt{NNSVD-LRC} function from \url{https://sites.google.com/site/nicolasgillis/code} \cite{atif_improved_2019}.
    \item \textbf{Multivariate empirical mode decomposition} was applied using code from \url{http://www.commsp.ee.ic.ac.uk/~mandic/research/emd.htm} \cite{rehman_multivariate_2010}.
\end{itemize}

For the remainder of the analysis and the construction of all figures we used Python version 3.5. Either standard functions obtained from published libraries supported by Python were used or custom code written in Python. The main functions used in the analysis are listed below:

\begin{itemize}
    \item \textbf{Hilbert transform}: \texttt{scipy.signal.hilbert}
    \item \textbf{LASSO}: \texttt{sklearn.linear\_model.Lasso}
    \item \textbf{k-fold cross-validation}: \texttt{sklearn.model\_selection.kFold}
    \item \textbf{Multiple Linear Regression}:
    \texttt{statsmodels.api.ols}
\end{itemize}

Our analysis code and data (post processing) can be found on \url{https://github.com/cnnp-lab/} (will be added).

\FloatBarrier

\section{Results\label{sec:results}}

We analysed fluctuations in band power for 18 subjects with focal epilepsy. We investigated if fluctuations on specific timescales were driven by particular iEEG frequency bands or spatially localised activity. We then explored if these temporal fluctuations were associated with how seizures change within subjects. 

\subsection{iEEG band power patterns fluctuate on different timescales\label{subsec:band_power_fluctuates}}

After extracting band power in the main frequency bands ($\delta, \theta, \alpha, \beta, \gamma$) in 30~s non-overlapping sliding windows for each iEEG channel (Fig.~\ref{fig:NMF_illustration}a,b), we performed dimensionality reduction using non-negative matrix factorisation (NMF) approach. NMF effectively grouped channels and frequency bands to form components that represent specific band power patterns. Weights for channels and frequency bands in each component are shown as columns in matrix $W$, Fig.~\ref{fig:NMF_illustration}c. The expression coefficients of these components at each time point was then given by the $H$ matrix, which essentially yielded a time series for each component (Fig.~\ref{fig:NMF_illustration}d). The weight represented a subject-specific pattern of EEG band power activity across channels, and the strength of expression of this pattern at any given time point was given by the expression coefficients. In short, the set of coefficient time series (rows in $H$) indicated the fluctuations of subject-specific EEG spectral patterns over time. 

\begin{figure}[H]
    \hspace{-0.8cm}
    \includegraphics[scale = 1]{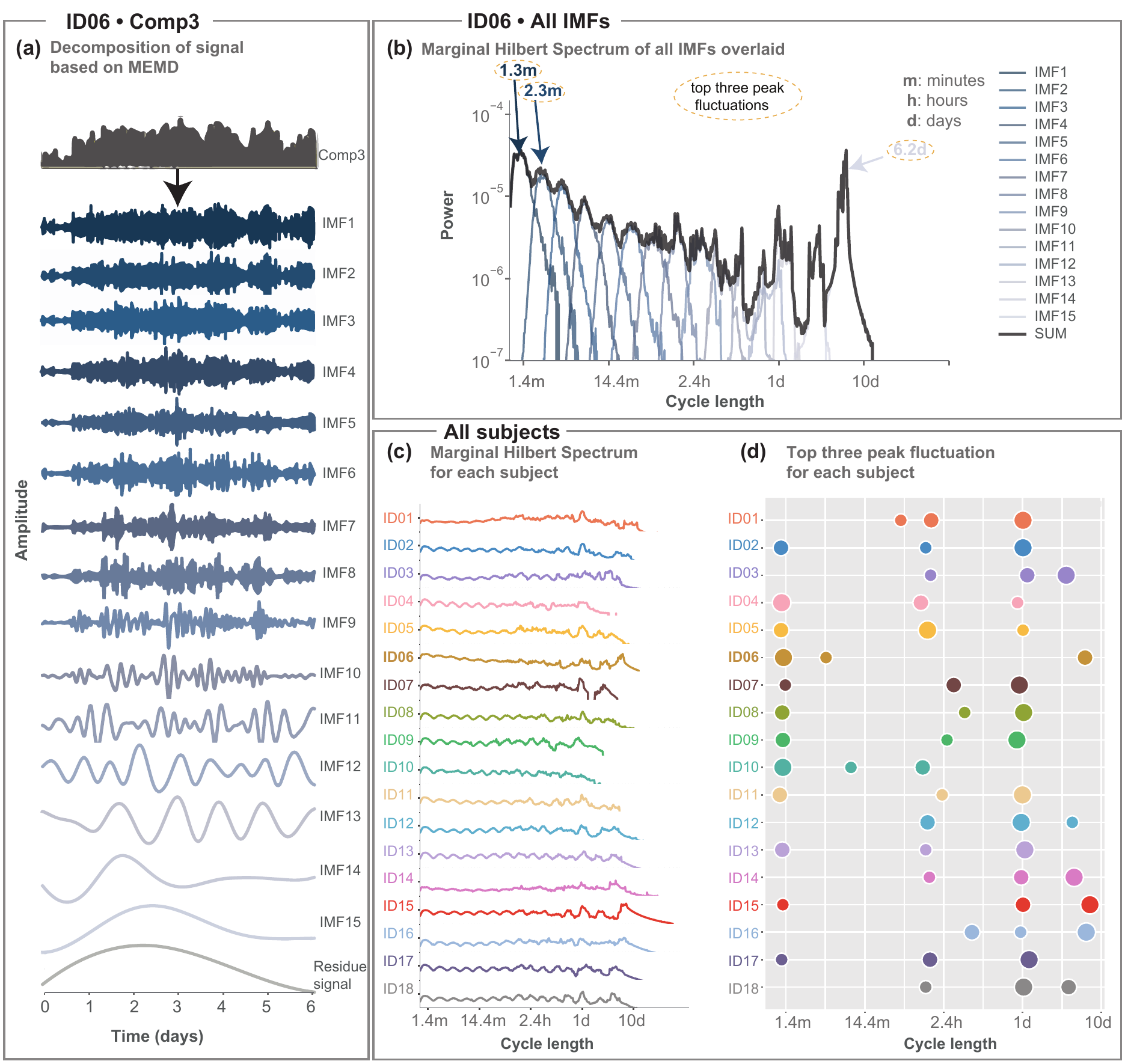}
       \caption{\textbf{MEMD detects fluctuations on different timescales for each subject.} (a) MEMD yields 16 IMFs in example subject ID06. Only one dimension of the IMF (corresponding to the first NMF component) is shown for simplicity. IMFs are presented in ascending order (fastest to slowest, top to bottom). The last trace is the residue signal. (b) Marginal Hilbert spectrum for all IMFs aggregated across all dimensions in example subject ID06. The black line represents the Marginal Hilbert frequency spectrum across all IMFs. The x-axis shows the instantaneous period length (inverse of instantaneous frequency), which we also termed `cycle length'. Top three peak fluctuations are indicated with arrows. (c) Marginal Hilbert frequency spectrum across all IMFs for each subject. (d) Bubble plot of peak fluctuations for the three highest power densities according to the Marginal Hilbert frequency spectrum across all IMFs for each subject. The size of the bubbles indicates the first, second and third peak in descending order.}
    \label{fig:MEMD_method}
\end{figure}

For each subject, we then used Multivariate Empirical Mode Decomposition (MEMD) to determine the different fluctuations on different timescales for each NMF coefficient time series. Figure~\ref{fig:MEMD_method}a shows the MEMD results for a single NMF component in example subject ID06, yielding 15 Intrinsic Mode Functions (IMFs) and a residue signal. Faster IMFs (e.g., IMF1, 2 and 3) are often thought to contain noise, but might also reflect genuine fluctuations in the initial signal, such as cyclic alternating pattern \cite{parrino_cyclic_2014}. For simplicity, we retained all IMFs for the subsequent main results and refer the reader to \ref{subsec:noisyIMF} for a more detailed analysis of noisy IMFs based on permutation test.

Using the instantaneous frequency and amplitude through the Hilbert transform, we obtained the marginal spectral densities of each IMF in each dimension. Figure~\ref{fig:MEMD_method}b shows the marginal spectral densities averaged across all dimensions for each IMF (blue lines) for example subject ID06. Some distinct peaks are seen especially in the slower IMFs, e.g. IMF13 (at cycle length of $\approx 1$ day ), IMF14 (at cycle length of $\approx 3.3$ days), IMF9 (cycle length $\approx 3$ hours), IMF8 (cycle length $\approx 1.6$ hours), etc. Note that both EMD and MEMD essentially act as dyadic filter banks \cite{wu_study_2004, flandrin_empirical_2004, ur_rehman_filter_2011}; thus, the dyadic pattern seen in the faster IMFs is not surprising. \ref{subsec:noisyIMF} shows the marginal spectral densities corrected for potential noise fluctuations.
    
As expected from previous literature \cite{baud_multi-day_2018, karoly_interictal_2016}, we found that all subjects displayed circadian band power fluctuations (Fig.~\ref{fig:MEMD_method}c). The presence of these circadian fluctuations helps validate our approach for extracting relevant timescales in interictal fluctuations. Meanwhile, fluctuations on other timescales were more subject-specific in cycle length. For 10 out of 18 subjects (ID01, ID02, ID07, ID08, ID09, ID11, ID12, ID13, ID17 and ID18) the circadian fluctuation had the highest density (Fig.~\ref{fig:MEMD_method}d). For six subjects (ID03, ID04, ID05, ID14, ID15 and ID16) the circadian fluctuation was slightly lower in density, as the highest density was seen in slower or faster IMFs. For two subjects (ID06 and ID10) the circadian fluctuation did not feature in the top three highest densities, but a peak at 1 cycle per day can still be observed in ID06 (Fig.~\ref{fig:MEMD_method}c).

\subsection{All iEEG frequency bands contribute to the circadian IMF\label{subsec:24h_IMF}}

\begin{figure}[H]
    \hspace{-0.8cm}
    \includegraphics[scale = 1]{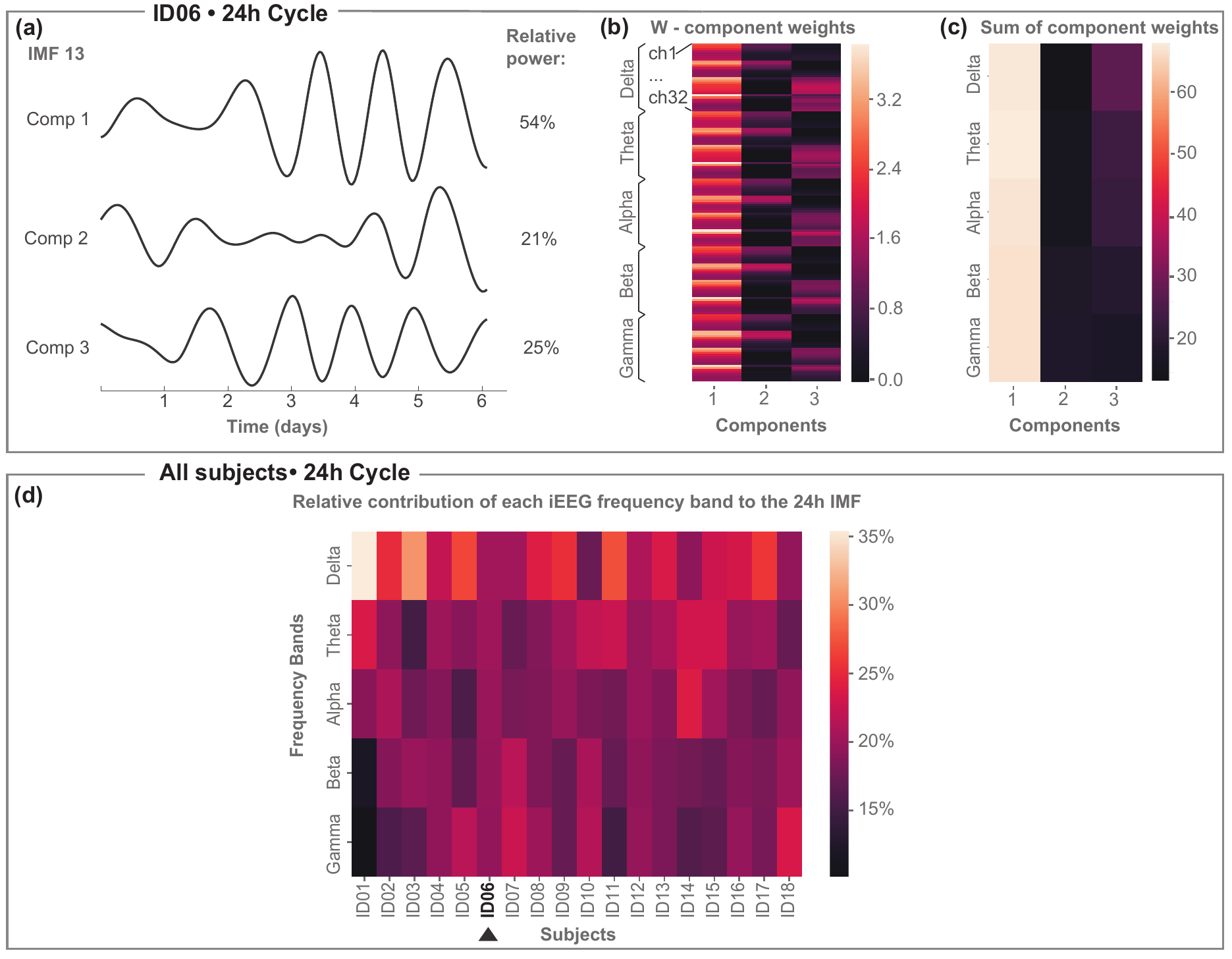}
       \caption{\textbf{Contribution of iEEG main frequency bands to the circadian IMF.} (a) IMF 13 in example subject ID06 shows circadian fluctuations across all three dimensions, each of which corresponds to an NMF component. Dimension 1 shows the highest relative power in this IMF. (b) W component weight matrix (same as Fig.~\ref{fig:NMF_illustration}c). (c) The sum of the component weights across all channels within each frequency band. (d) Contribution of each iEEG frequency band to the circadian IMF across all subjects obtained by forming the sum over the matrix in (c) weighted by the relative power in (a). To be able to compare subjects to each other, each column here has been normalised to form a percentage contribution.}
    \label{fig:W_connection}
\end{figure}

Following the observation of a circadian fluctuation in all subjects, we assessed the contribution of each iEEG frequency band to the circadian IMF. We first determined the circadian IMF, which was IMF 13 in example subject ID06  (Fig.~\ref{fig:W_connection}a). We then calculated the relative power in each dimension of the IMF, each of which corresponded to an NMF component. For example, in subject ID06, the majority of its power (54\%) was concentrated in dimension 1 (Fig.~\ref{fig:W_connection}a). We also noted that the circadian fluctuation did not follow the same phase in all dimensions of the IMF, potentially indicating the presence of multiple processes fluctuating on a circadian timescale. Since we are interested in the overall contribution of each frequency band to the circadian cycle, we decided to assess the contribution of different frequency bands over all dimensions next.

From the dimensionality reduction step, we had already obtained the weights across all iEEG frequency bands and channels (matrix W, see Fig.~\ref{fig:W_connection}b). For each NMF component, we computed the weight of each frequency band by summing the weights of that frequency band across all channels (Fig.~\ref{fig:W_connection}c). Finally, a sum weighted by the relative power in the IMF over all dimensions was obtained representing the relative contribution of each frequency band to the IMF. For most subjects, $\delta$ band power contribution was slightly higher compared to the other frequency bands for the circadian IMF. However, other frequency bands also contributed to the circadian IMF in most subjects (Fig.~\ref{fig:W_connection}d). 

\subsection{Subsets of channels contribute to infradian band power fluctuations\label{subsec:IMF_slower_than_24h}}

\begin{figure}[hbp]
    \centering
    \includegraphics[scale = 1]{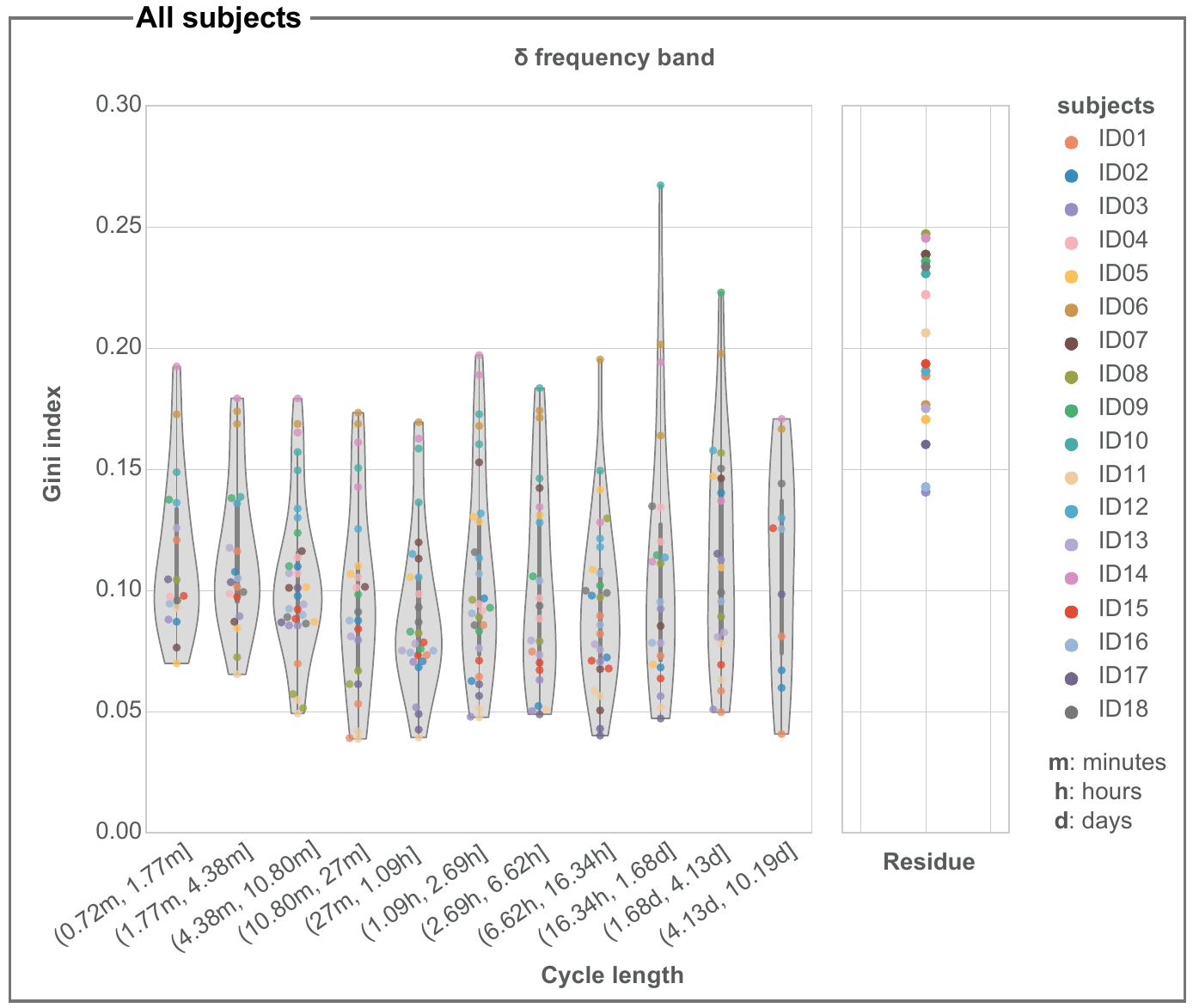}
       \caption{\textbf{Gini index of IMFs for the $\delta$ frequency band across all subjects.} 
       Across all subjects, we grouped IMFs based on their peak IMF cycle length and show the distribution of the corresponding Gini indices as a violin plot with enclosed box plot. The thick grey bar represents the inter-quartile range. For visualisation, we converted the peak frequency to cycle length (x-axis). The residue is shown separately.}
    \label{fig:Gini_index}
\end{figure}

Within each frequency band we also investigated the contribution of each channel to an IMF. Specifically, we investigated if the contributions were heterogeneous across channels. We used the 
Gini index as a measure of spatial heterogeneity, where 0 (1) indicates a completely homogeneous (heterogeneous) channel contribution for each IMF.
Figure~\ref{fig:Gini_index} shows the distribution of Gini indices of all IMFs in the $\delta$ band across all subjects, where IMFs are grouped by the IMF peak frequency. Results for other iEEG main frequency bands are similar and shown in Supporting Fig.~\ref{fig:Gini_index_rest}. Overall, the Gini indices are low for all IMFs, indicating that IMFs are not driven by a small group of channels. However, there is a clear tendency for long-term trends to display a higher Gini index, indicating that a subsets of channels may contribute more to those.

\subsection{Band power IMF fluctuations are associated with seizure dissimilarity in most subjects\label{subsec:IMF_explain_sz_diss}}

As the final part of our analysis, we investigated if these fluctuations on different timescales influenced, or modulated, changes in seizure evolutions over time in individual subjects. Particularly, we previously showed that seizure network evolutions change over time in every subject, and that these changes could be explained by hypothetical circadian or longer timescale modulators \cite{schroeder_seizure_2020}. Hence, we explored if the subject-specific fluctuations represented by the IMFs were associated with changes in seizure evolutions.

\begin{figure}[tp!]
    \hspace{-0.8cm}
    \includegraphics[scale = 1]{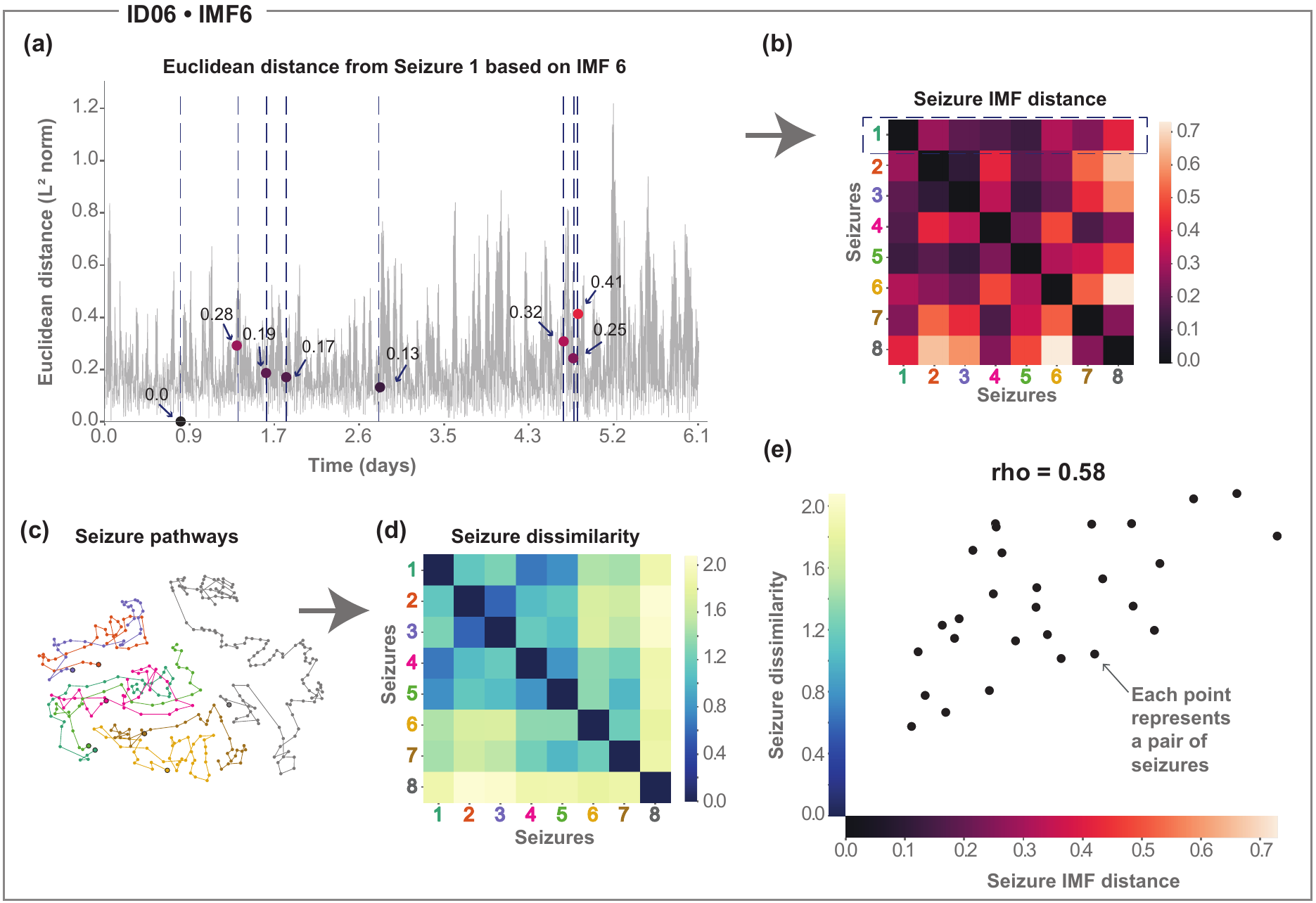}
     \caption{\textbf{Relating seizure dissimilarity and IMF seizure distance.} Throughout the figure we use example subject ID06 and IMF6. (a) Euclidean distance of all time points to the first seizure in terms of IMF6. Blue dashed vertical lines indicate seizure timing. Dots mark the value of the IMF distance to the first seizure and colours of dots correspond to the colour scale in (b). (b) Seizure IMF distance matrix for IMF6. The first row is a representation of the data in (a). (c) Visualising seizure evolutions as pathways (see \ref{suppl:visualDissMat} and \cite{schroeder_seizure_2020}). Seizures are displayed with distinct colours to distinguish seizure events. The starting points of seizures are marked with a black outline circle. In this projection, parts of seizures with similar network evolutions tend to be placed closer together, and seizures with similar evolutions will therefore approximately overlap (e.g., orange and purple pathways).  (d) Seizure dissimilarity matrix, capturing the differences in seizure evolutions over time between each pair of seizures. (e) Scatter plot of seizure dissimilarity and the seizure IMF distance (Spearman's correlation, $\rho = 0.58$).}
     \label{fig:mantel}
\end{figure}

For each IMF in each subject, we first determined their corresponding seizure IMF Euclidean distance matrix (Fig.\ref{fig:mantel}a,b). For example, in subject ID06's IMF6, we calculated the Euclidean distance of every time point to the time point of the first seizure (Fig.\ref{fig:mantel}a) across all dimensions. By reading out all the Euclidean distances to all the other seizure time points, we obtained the first row of the seizure IMF Euclidean distance matrix (Fig.\ref{fig:mantel}b). The same process was repeated for all other seizures in this subject. This distance matrix had dimensions of number of seizures by number of seizures and represented how different the IMF state was for each seizure pair.

By using the same techniques as in \cite{schroeder_seizure_2020}, we obtained a seizure dissimilarity matrix, which expressed the dissimilarity of each pair of seizure evolutions through the space of network dynamics (Fig.\ref{fig:mantel}c,d). The seizure dissimilarity matrix thus quantified how much each pair of seizures differed within a subject. By relating the set of seizure dissimilarities to the corresponding set of IMF Euclidean distance, we investigated if there was an association between changes in seizure evolutions and interictal band power fluctuations (Fig.\ref{fig:mantel}e).

\begin{figure}[hbp]
    \hspace{-0.8cm}
    \includegraphics[scale = 1]{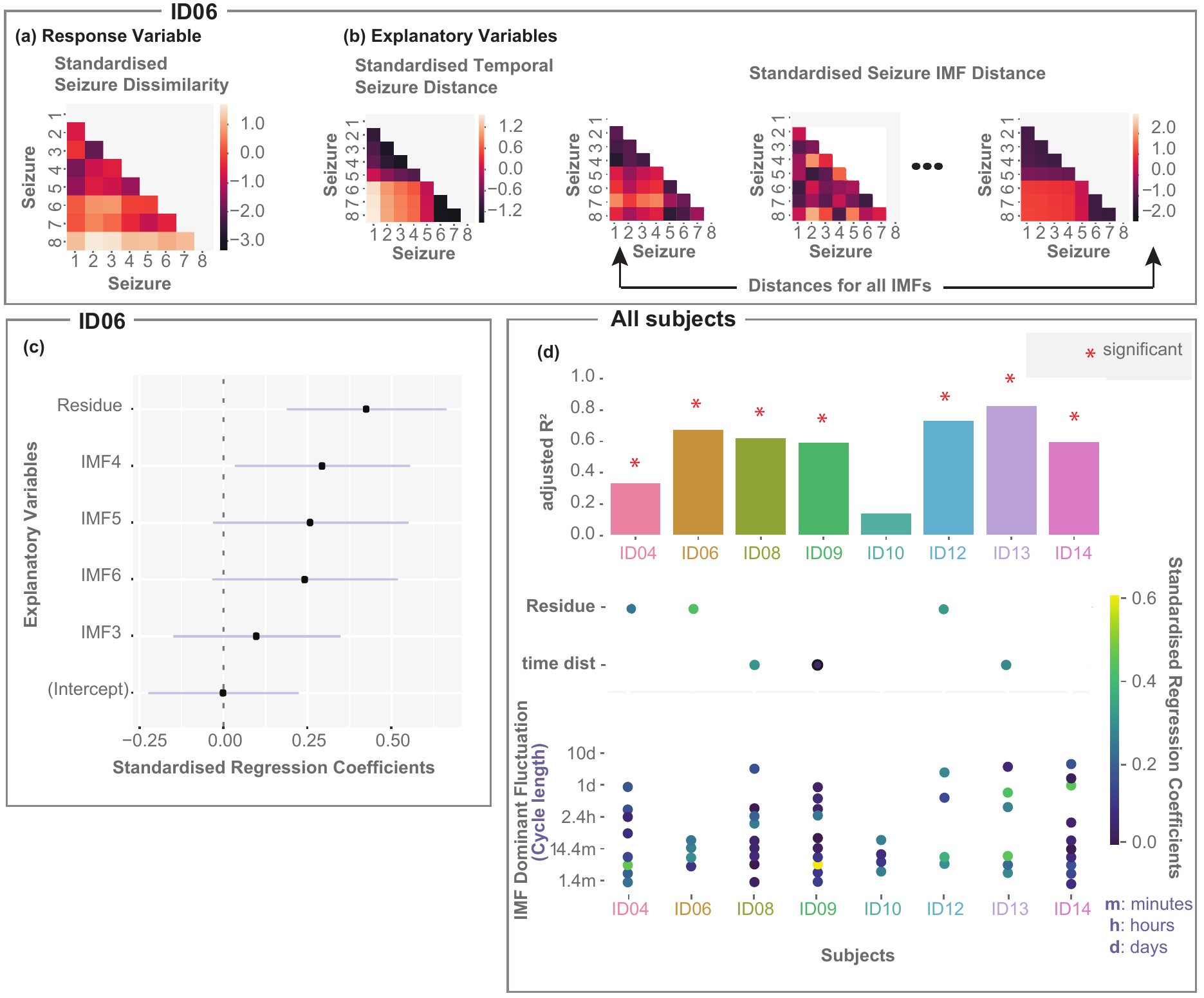}
\
    \caption{\textbf{A combination of IMF seizure distances on different timescales can explain seizure dissimilarity in most subjects in a multiple regression model.} 
    (a) Standardised seizure dissimilarity matrix (response variable). Only the lower triangle of the symmetric matrix is shown, where each entry serves as an observation. (b) Explanatory variables: The matrix on the left shows the standardised temporal seizure distance. Each entry corresponds to the absolute time difference between seizures. The remaining matrices are standardised seizure IMF distance matrices. 
    (c) Coefficient estimates (black dots), based on ordinary least squares regression for subject ID06, with lines indicating $95\%$ confidence intervals. Only five explanatory variables were left after performing variable selection based on constrained LASSO. (d) Summary across subjects based on Ordinary Least Squares (OLS) models with explanatory variables obtained by the constrained LASSO. Top: Bar chart of the adjusted $R^2$. Red stars indicate p-values $ \le0.05$. 
    Bottom: Scatter plot indicating the OLS coefficient estimates for the residue, temporal distance (when these variables remained in the model), together with explanatory IMFs and their corresponding IMF peak frequency for each subject. For visualisation, we converted the peak frequency to cycle length.}
    \label{fig:lasso}
\end{figure}

To generalise this approach to all IMFs in a subject, we fitted a multiple linear regression model, where the sets of seizure IMF distances (derived from different IMFs) were explanatory variables and the seizure dissimilarity was the response variable (Fig. \ref{fig:lasso}a). We also included the EMD residue signal and temporal distance between seizures (i.e. how far apart in time each seizure pair occurred) as explanatory variables to model fluctuations of longer timescales than the recording time. The observations were pairs of seizures. After LASSO variable selection and linear regression, the estimated regression coefficients for example subject ID06 are shown in Fig. \ref{fig:lasso}(c). For this particular subject, the strongest explanatory effect (as measured by the standardised regression coefficients, also know as beta-weights) was seen in the EMD residue signal followed by some faster IMFs (IMF $[\textbf{Cycle length}]$: IMF3 $[\textbf{4 min}]$, IMF4 $[\textbf{7.5 min}]$, IMF5 $[\textbf{15 min}]$ and IMF6 $[\textbf{26 min}]$). According to the model, $67.42\%$ of the variability in seizure dissimilarities was explained by explanatory variables (i.e. adjusted $R^2 = 0.6742$).

Across subjects, we fitted the multiple linear regression model only for subjects with at least six seizures, resulting in eight subjects with analysed seizure evolutions. Out of our cohort of eight subjects, six had an adjusted $R^2$ around or above 0.6 (Fig.\ref{fig:lasso}d). Fig.~\ref{fig:adjusted_Rsquared} additionally shows that the adjusted $R^2$ values would have not occurred by chance in any subject except for ID10. For 6 out of 8 subjects, circadian IMFs were also part of the explanatory variables (Fig.\ref{fig:lasso}d). Ultradian IMFs also tended to remain as explanatory variables in the models for all subjects. Temporal distance between seizures remained as an explanatory variable in three subjects, and the residue signal also remained as an explanatory variable in three additional subjects.  Overall, a subject-specific combination of different fluctuations were provided a good explanation of seizure variability in most subjects.

Note that band power fluctuations are not expected to trivially correlate with how seizures change, as (i) the seizure network evolutions changes are detected on a finer timescale (seconds) using a functional network measure of the time series rather than a spectral property; (ii) seizure onset network patterns (as measured by functional networks) are also expected to differ substantially from pre-ictal network patterns \cite{shah_characterizing_2019}; (iii) the impact of seizures on the band power fluctuation are most likely to be limited to one or few 30~s windows and hence also likely to be limited to the fastest IMF only. In \ref{suppl:Corr_band_power} we show that the band power without being decomposed into different timescales does not explain how seizures change, indicating that our results did not arise from trivial associations between seizure evolutions and their corresponding interictal periods. In \ref{subsec:raw_analysis_onset-1} we also reproduced our results using the pre-ictal (one 30~s window ahead of the seizure) band power fluctuations, which were not impacted by seizure evolutions.

\newpage
\FloatBarrier

\section{Discussion\label{sec:discussion}}

We analysed fluctuations in subject-specific iEEG band power patterns over time and found that these patterns fluctuate over a wide range of timescales (from minutes to days), including a strong circadian fluctuation in most patients. A subject-specific combination of these fluctuations provided a good explanation (adjusted $R^2>=0.6$) for how seizure EEG spatio-temporal evolutions change from one seizure to the next within the same subject. Based on these findings, we suggest that band power fluctuations in continuously recorded EEG may be a marker of modulators of seizure activity.
 
Fluctuations on various timescales of the continuous EEG have been reported in several studies using iEEG recordings. The prevalence of a strong circadian rhythm in EEG patterns has long been known \cite{scheich_interval_1969, spencer_circadian_2016, smyk_circadian_2020, aeschbach_two_1999, cummings_diurnal_2000}. Weaker ultradian (more than 1 cycle per day) rhythms have been reported in long-term EEG band power \cite{kaiser_ultradian_2008, chapotot_high_2000} and functional connectivity  \cite{mitsis_functional_2020}. Subject-specific multidien (multi-day, i.e. less than 1 cycle per day) rhythms have also been detected in e.g. the rate of interictal epileptiform activity \cite{baud_multi-day_2018,karoly_interictal_2016}, and the variance and autocorrelation of EEG signals \cite{maturana_critical_2020}. In agreement, we observed the circadian cycle in all subjects and additional fluctuations on ultradian and multidien timescales that were subject-specific.

These fluctuations of EEG features on different timescales most likely reflect biological processes. However, the mapping from EEG biomarkers to underlying time-varying processes is incomplete. Various hypotheses exist regarding the interpretation of these EEG fluctuations \cite{bernard_circadianmultidien_2021,karoly_cycles_2021,rao_cues_2021}, and their possible drivers include hormonal and metabolic cycles, changes in antiepileptic medications, and external influences such as the weather \cite{badawy_epilepsy_2012, meisel_intrinsic_2015, karoly_cycles_2021}. In this work, we therefore took a subject-specific data-driven approach that allowed us to detect any prominent fluctuations, regardless of their subject-specific source. Future work will explore a wider range of EEG biomarkers and elucidate the exact mapping between different fluctuations and the underlying physiological or pathological processes.

Additionally, we make two observations about band power fluctuations on different timescales. First, we saw that different frequency bands appeared to contribute a similar amount to the circadian fluctuation of iEEG band power, although subtle subject-specific patterns of contribution are also noted. However, our analysis was performed across all dimensions of our data. The different dimensions of the IMF can display phase and amplitude differences (see e.g. Fig.~\ref{fig:W_connection}a), indicating that different circadian fluctuations (with different phases) exist in each subject, as has been reported before \cite{aeschbach_two_1999}. Future work may wish to investigate the frequency contributions to different dimensions of IMFs and also relate those IMFs to other physiological variables such as body temperature or plasma melatonin \cite{aeschbach_two_1999}.

The second observation is that slower (multi-day fluctuations, and slow trends)  tended result from changes in subsets of channels, whereas faster (circadian and ultradian) fluctuations tended to arise as a more equal contribution from all channels. A limitation in our analysis is that iEEG provides limited spatial coverage and the electrode layout is patient-specific, making it difficult to compare patterns of band power fluctuations across subjects. To fully uncover the spatial and frequency band contributions to each dimension of each IMF, we suggest that future work should consider the spatial location of iEEG channels and perform an iterative combination of dimensionality reduction and empirical mode decomposition to find components and their contributions for each IMF. From a clinical perspective, information on the spatial coverage and location of the electrodes would further allow us to investigate the overlap of the location of these temporal fluctuations with the epileptogenic zone in focal epilepsies.

We applied empirical mode decomposition (EMD) to derive band power fluctuations on different timescales. EMD is a popular data-driven adaptive method with applications on broad range of scientific topics, such as geology \cite{battista_application_2007}, hydrology \cite{hu_soil_2013}, and neuroscience \cite{huang_application_2013, rojas_application_2013} amongst many others. It is suitable for extracting fluctuations on different timescales without assumptions of local stationarity, linearity, or specific basis functions, and for these reasons preferable for our application. Since EMD does not require a basis function to identify different timescales of fluctuations, it also does not generate harmonics (as in Fourier or Wavelet-type approaches) of fluctuations, making the decomposed cycles easier to interpret. However, EMD also has some limitations. Most notably, the IMFs' timescales of fluctuations may overlap, which is known as `mode mixing' \cite{ur_rehman_filter_2011}. EMD may also struggle to distinguish two distinct fluctuations that have very similar periods, and they may be merged into one IMF. Ongoing developments \cite{xue_adaptively_2015, deering_use_2005,li_improved_2015} in this area may overcome these limitations. Future work should explore how to capture non-stationary \cite{kaplan_nonstationary_2005}, non-linear \cite{stam_nonlinear_2005}, and potentially hierarchical \cite{vidaurre_brain_2017} time-varying properties of the continuously recorded EEG.

Our main goal was to investigate if there is an association between variability in seizure evolutions and fluctuations in long-term iEEG band power. Changes in seizure evolutions can be quantitatively described using seizure dissimilarity, which captures how different any pair of seizures are in a given subject in terms of their seizure network evolutions \cite{schroeder_seizure_2020}. Previous work has also shown that fluctuations in seizure evolutions were well-explained by processes incorporating Gaussian noise, circadian, and/or slower timescales of changes in most subjects \cite{schroeder_seizure_2020}. In agreement with this work, we found that circadian or multidien fluctuations contributed strongly in most subjects in explaining seizure dissimilarity. In three subjects (ID04, ID06, ID12), the residue signal also contributed to the explanation, indicating that fluctuations on longer periods than the recording durations also played a role. Interestingly, we also found many faster (ultradian) fluctuations as explanatory variables in most subjects. These fluctuations could be contributing explanatory power through what previously was modelled as noise \cite{schroeder_seizure_2020}. However, there may also be a true biological fluctuation underpinning the explanation; faster fluctuations in the EEG have also been reported e.g. in the cyclic alternating pattern \cite{parrino_cyclic_2014}. With larger datasets using more seizures recorded over a longer period, future work should investigate ultradian contributions carefully and assess if noise would perform as well as the cumulative ultradian contributions.

While fluctuations in long-term iEEG band power can explain seizure dissimilarity fairly well, this association should not be interpreted as causal evidence. The observed band power fluctuations can be understood as signatures of multiple biological processes, which could directly dictate seizure evolutions or be co-modulated by the same upstream processes as the seizure evolutions. Our data cannot distinguish these cases. Additional fluctuations that are not captured by iEEG band power may also explain changes in seizure evolutions, and a more detailed analysis of the exact fluctuations and the differences in specific seizure features may be more informative. Interestingly, band power fluctuations did not account for all the seizure variability in most subjects. The highest adjusted $R^2$ was around 0.8 and the unexplained variability based on the models suggests that there are additional factors, or possibly a level of stochasticity, that impact seizure evolutions. Nevertheless, to make our findings clinically useful, e.g. as a predictive model of upcoming seizure evolutions or seizure severity, neither causality nor completeness of the predictors is required. Our results indicate that a predictive model of seizure evolutions is possible with continuously recorded features such as iEEG band power, and this model should achieve good predictive performance in the majority of subjects.

To improve predictive performance, other factors could be considered in future, e.g. the anti-epileptic drug (AED) level at any given time, or additional EEG features. Specifically, it is well-known that AED changes and withdrawal can change the severity and evolutions of seizures. For example, bilateral tonic-clonic seizures are more prevalent when AED levels are reduced \cite{pensel_predictors_2020}. AEDs have further been shown to impact inter- and peri-ictal brain activity \cite{badawy_peri-ictal_2009,meisel_intrinsic_2015}, making it an important feature to consider. We were unable to include this information in the current study, but future studies may wish to investigate how AED levels impact iEEG band power \cite{arzy_antiepileptic_2010}, in combination with their potential explanatory power for seizure evolution changes.  

Any multi-way association between continuously recorded brain activity, seizure evolutions, and treatments (such as AEDs) has the potential to introduce entirely new treatment strategies. If, for example, particular interictal EEG signatures predict more severe seizures, and these signatures are also influenced by AED dose, then one can hypothesise that responsively adapting AED dose according to these interictal signatures might decrease seizure severity. If the hypothesis can be verified, then on-demand drug-delivery systems programmed to respond to patient-specific interictal signatures could become the next generation of epilepsy treatments \cite{carney_chronotherapy_2014, manganaro_need_2017,ramgopal_chronopharmacology_2013}.

In a more general context, our work is another contribution to the wider literature of explaining ictal features from interictal EEG features or hypothesised circadian/multidien rhythms. For example, studies have established that there is often a subject-specific relationship between fluctuations of interictal EEG features and the timing of ictal events \cite{baud_multi-day_2018,karoly_interictal_2016,mitsis_functional_2020, maturana_critical_2020, leguia_seizure_2021}. Interestingly, we found no evidence of an association between band power fluctuations of the interictal EEG and seizure occurrence (data not shown). Seizures were not more likely to occur during particular phases of particular IMFs in most subjects in our data set. This finding is in agreement with a previous study \cite{mitsis_functional_2020} that reported functional network fluctuations, rather than band power fluctuations, to be more predictive of seizure timing. Future work should investigate temporal fluctuations in a range of EEG features, such as band power \cite{cummings_diurnal_2000}, functional connectivity \cite{mitsis_functional_2020}, high frequency oscillations \cite{gliske_variability_2018}, variance, and autocorrelation \cite{maturana_critical_2020}. Apart from seizure timing, our work has shown that band power fluctuations on different timescales also explain changes in seizure evolutions. Future work should explore this avenue further to illuminate the exact processes and timescales that modulate or dictate the various aspects of a seizure.

Fluctuating interictal EEG features may not only correlate with clinical seizure timing, but also with seizure evolutions on infradian, circadian, and ultradian timescales. In the future, it may be possible to use these temporal patterns of EEG fluctuations to predict seizure evolutions. Prediction of various seizure features, including seizure evolution and seizure severity is a critical unmet need for people with epilepsy. If successful, it would open up new opportunities for therapeutics and maximising quality of life. 

\section*{Acknowledgements}
\subsection*{Funding Statement}

MP was supported by the Engineering and Physical Sciences Research Council, Centre for Doctoral Training in Cloud Computing for Big Data (grant number EP/L015358/1).

Y.W. gratefully acknowledges funding from Wellcome Trust (208940/Z/17/Z). P.N.T. is supported by a UKRI Future Leaders Fellowship (MR/T04294X/1).

\subsection*{Conflict of Interest Disclosure}

The authors declare no competing interests.

\subsection*{Ethics Approval Statement}

The project was granted its approval by the Newcastle University Ethics Committee (Ref: 18818/2019).

\subsection*{Patient Consent Statement }

All subjects formally consented to their iEEG data being used for research purposes \cite{burrello_laelaps:_2019}.

\FloatBarrier
\bibliography{main.bib} 

\begin{thebibliography}{10}

\bibitem{fisher_ilae_2014}
R.~S. Fisher, C.~Acevedo, A.~Arzimanoglou, A.~Bogacz, J.~H. Cross, C.~E. Elger,
  J.~Engel, L.~Forsgren, J.~A. French, M.~Glynn, D.~C. Hesdorffer, B.~I. Lee,
  G.~W. Mathern, S.~L. Moshé, E.~Perucca, I.~E. Scheffer, T.~Tomson,
  M.~Watanabe, and S.~Wiebe, ``{ILAE} {Official} {Report}: {A} practical
  clinical definition of epilepsy,'' {\em Epilepsia}, vol.~55, no.~4,
  pp.~475--482, 2014.

\bibitem{chen_treatment_2018}
Z.~Chen, M.~J. Brodie, D.~Liew, and P.~Kwan, ``Treatment {Outcomes} in
  {Patients} {With} {Newly} {Diagnosed} {Epilepsy} {Treated} {With}
  {Established} and {New} {Antiepileptic} {Drugs}: {A} 30-{Year} {Longitudinal}
  {Cohort} {Study},'' {\em JAMA neurology}, vol.~75, pp.~279--286, Mar. 2018.

\bibitem{kramer_coalescence_2010}
M.~A. Kramer, U.~T. Eden, E.~D. Kolaczyk, R.~Zepeda, E.~N. Eskandar, and S.~S.
  Cash, ``Coalescence and {Fragmentation} of {Cortical} {Networks} during
  {Focal} {Seizures},'' {\em Journal of Neuroscience}, vol.~30,
  pp.~10076--10085, July 2010.

\bibitem{schindler_forbidden_2011}
K.~Schindler, H.~Gast, L.~Stieglitz, A.~Stibal, M.~Hauf, R.~Wiest, L.~Mariani,
  and C.~Rummel, ``Forbidden ordinal patterns of periictal intracranial {EEG}
  indicate deterministic dynamics in human epileptic seizures,'' {\em
  Epilepsia}, vol.~52, no.~10, pp.~1771--1780, 2011.

\bibitem{schevon_evidence_2012}
C.~A. Schevon, S.~A. Weiss, G.~McKhann, R.~R. Goodman, R.~Yuste, R.~G. Emerson,
  and A.~J. Trevelyan, ``Evidence of an inhibitory restraint of seizure
  activity in humans,'' {\em Nature Communications}, vol.~3, p.~1060, Sept.
  2012.

\bibitem{burns_network_2014}
S.~P. Burns, S.~Santaniello, R.~B. Yaffe, C.~C. Jouny, N.~E. Crone, G.~K.
  Bergey, W.~S. Anderson, and S.~V. Sarma, ``Network dynamics of the brain and
  influence of the epileptic seizure onset zone,'' {\em Proceedings of the
  National Academy of Sciences}, vol.~111, pp.~E5321--E5330, Dec. 2014.

\bibitem{wagner_microscale_2015}
F.~B. Wagner, E.~N. Eskandar, G.~R. Cosgrove, J.~R. Madsen, A.~S. Blum, N.~S.
  Potter, L.~R. Hochberg, S.~S. Cash, and W.~Truccolo, ``Microscale
  spatiotemporal dynamics during neocortical propagation of human focal
  seizures,'' {\em NeuroImage}, vol.~122, pp.~114--130, Nov. 2015.

\bibitem{truccolo_single-neuron_2011}
W.~Truccolo, J.~A. Donoghue, L.~R. Hochberg, E.~N. Eskandar, J.~R. Madsen,
  W.~S. Anderson, E.~N. Brown, E.~Halgren, and S.~S. Cash, ``Single-neuron
  dynamics in human focal epilepsy,'' {\em Nature Neuroscience}, vol.~14,
  pp.~635--641, May 2011.

\bibitem{cook_human_2016}
M.~J. Cook, P.~J. Karoly, D.~R. Freestone, D.~Himes, K.~Leyde, S.~Berkovic,
  T.~O'Brien, D.~B. Grayden, and R.~Boston, ``Human focal seizures are
  characterized by populations of fixed duration and interval,'' {\em
  Epilepsia}, vol.~57, no.~3, pp.~359--368, 2016.

\bibitem{marciani_effects_1986}
M.~G. Marciani and J.~Gotman, ``Effects of {Drug} {Withdrawal} on {Location} of
  {Seizure} {Onset},'' {\em Epilepsia}, vol.~27, no.~4, pp.~423--431, 1986.

\bibitem{karthick_prediction_2018}
P.~A. Karthick, H.~Tanaka, H.~M. Khoo, and J.~Gotman, ``Prediction of secondary
  generalization from a focal onset seizure in intracerebral {EEG},'' {\em
  Clinical Neurophysiology}, vol.~129, pp.~1030--1040, May 2018.

\bibitem{naftulin_ictal_2018}
J.~S. Naftulin, O.~J. Ahmed, G.~Piantoni, J.-B. Eichenlaub, L.-E. Martinet,
  M.~A. Kramer, and S.~S. Cash, ``Ictal and {Preictal} {Power} {Changes}
  {Outside} of the {Seizure} {Focus} {Correlate} with {Seizure}
  {Generalization},'' {\em Epilepsia}, vol.~59, p.~1398, July 2018.

\bibitem{pensel_predictors_2020}
M.~C. Pensel, M.~Schnuerch, C.~E. Elger, and R.~Surges, ``Predictors of focal
  to bilateral tonic-clonic seizures during long-term video-{EEG} monitoring,''
  {\em Epilepsia}, vol.~61, no.~3, pp.~489--497, 2020.

\bibitem{alarcon_power_1995}
G.~Alarcon, C.~D. Binnie, R.~D.~C. Elwes, and C.~E. Polkey, ``Power spectrum
  and intracranial {EEG} patterns at seizure onset in partial epilepsy,'' {\em
  Electroencephalography and Clinical Neurophysiology}, vol.~94, pp.~326--337,
  May 1995.

\bibitem{schroeder_seizure_2020}
G.~M. Schroeder, B.~Diehl, F.~A. Chowdhury, J.~S. Duncan, J.~d. Tisi, A.~J.
  Trevelyan, R.~Forsyth, A.~Jackson, P.~N. Taylor, and Y.~Wang, ``Seizure
  pathways change on circadian and slower timescales in individual patients
  with focal epilepsy,'' {\em Proceedings of the National Academy of Sciences},
  May 2020.

\bibitem{saggio_taxonomy_2020}
M.~L. Saggio, D.~Crisp, J.~M. Scott, P.~Karoly, L.~Kuhlmann, M.~Nakatani,
  T.~Murai, M.~Dümpelmann, A.~Schulze-Bonhage, A.~Ikeda, M.~Cook, S.~V.
  Gliske, J.~Lin, C.~Bernard, V.~Jirsa, and W.~C. Stacey, ``A taxonomy of
  seizure dynamotypes,'' {\em eLife}, vol.~9, July 2020.

\bibitem{jobst_secondarily_2001}
B.~C. Jobst, P.~D. Williamson, T.~B. Neuschwander, T.~M. Darcey, V.~M. Thadani,
  and D.~W. Roberts, ``Secondarily {Generalized} {Seizures} in {Mesial}
  {Temporal} {Epilepsy}: {Clinical} {Characteristics}, {Lateralizing} {Signs},
  and {Association} {With} {Sleep}–{Wake} {Cycle},'' {\em Epilepsia},
  vol.~42, no.~10, pp.~1279--1287, 2001.

\bibitem{jin_prevalence_2017}
B.~Jin, S.~Wang, L.~Yang, C.~Shen, Y.~Ding, Y.~Guo, Z.~Wang, J.~Zhu, S.~Wang,
  and M.~Ding, ``Prevalence and predictors of subclinical seizures during scalp
  video-{EEG} monitoring in patients with epilepsy,'' {\em International
  Journal of Neuroscience}, vol.~127, pp.~651--658, Aug. 2017.

\bibitem{sunderam_epileptic_2007}
S.~Sunderam, I.~Osorio, and M.~G. Frei, ``Epileptic seizures are temporally
  interdependent under certain conditions,'' {\em Epilepsy Research}, vol.~76,
  pp.~77--84, Sept. 2007.

\bibitem{oken_short-term_1988}
B.~Oken and K.~Chiappa, ``Short-term variability in {EEG} frequency analysis,''
  {\em Electroencephalography and Clinical Neurophysiology}, vol.~69,
  pp.~191--198, Mar. 1988.

\bibitem{aeschbach_two_1999}
D.~Aeschbach, J.~R. Matthews, T.~T. Postolache, M.~A. Jackson, H.~A. Giesen,
  and T.~A. Wehr, ``Two circadian rhythms in the human electroencephalogram
  during wakefulness,'' {\em American Journal of Physiology-Regulatory,
  Integrative and Comparative Physiology}, vol.~277, pp.~R1771--R1779, Dec.
  1999.

\bibitem{geier_time-dependent_2015}
C.~Geier, K.~Lehnertz, and S.~Bialonski, ``Time-dependent degree-degree
  correlations in epileptic brain networks: from assortative to dissortative
  mixing,'' {\em Frontiers in Human Neuroscience}, vol.~9, p.~462, 2015.

\bibitem{geier_long-term_2017}
C.~Geier and K.~Lehnertz, ``Long-term variability of importance of brain
  regions in evolving epileptic brain networks,'' {\em Chaos: An
  Interdisciplinary Journal of Nonlinear Science}, vol.~27, p.~043112, Apr.
  2017.

\bibitem{mitsis_functional_2020}
G.~D. Mitsis, M.~N. Anastasiadou, M.~Christodoulakis, E.~S. Papathanasiou,
  S.~S. Papacostas, and A.~Hadjipapas, ``Functional brain networks of patients
  with epilepsy exhibit pronounced multiscale periodicities, which correlate
  with seizure onset,'' {\em Human Brain Mapping}, vol.~41, no.~8,
  pp.~2059--2076, 2020.

\bibitem{gliske_variability_2018}
S.~V. Gliske, Z.~T. Irwin, C.~Chestek, G.~L. Hegeman, B.~Brinkmann, O.~Sagher,
  H.~J.~L. Garton, G.~A. Worrell, and W.~C. Stacey, ``Variability in the
  location of high frequency oscillations during prolonged intracranial {EEG}
  recordings,'' {\em Nature Communications}, vol.~9, p.~2155, Dec. 2018.

\bibitem{karoly_interictal_2016}
P.~J. Karoly, D.~R. Freestone, R.~Boston, D.~B. Grayden, D.~Himes, K.~Leyde,
  U.~Seneviratne, S.~Berkovic, T.~O’Brien, and M.~J. Cook, ``Interictal
  spikes and epileptic seizures: their relationship and underlying
  rhythmicity,'' {\em Brain}, vol.~139, pp.~1066--1078, Apr. 2016.

\bibitem{conrad_spatial_2020}
E.~C. Conrad, S.~B. Tomlinson, J.~N. Wong, K.~F. Oechsel, R.~T. Shinohara,
  B.~Litt, K.~A. Davis, and E.~D. Marsh, ``Spatial distribution of interictal
  spikes fluctuates over time and localizes seizure onset,'' {\em Brain},
  vol.~143, pp.~554--569, Feb. 2020.

\bibitem{baud_multi-day_2018}
M.~O. Baud, J.~K. Kleen, E.~A. Mirro, J.~C. Andrechak, D.~King-Stephens, E.~F.
  Chang, and V.~R. Rao, ``Multi-day rhythms modulate seizure risk in
  epilepsy,'' {\em Nature Communications}, vol.~9, p.~88, Jan. 2018.

\bibitem{chen_spatiotemporal_2020}
Z.~Chen, D.~B. Grayden, A.~N. Burkitt, U.~Seneviratne, W.~J. D'Souza,
  C.~French, P.~J. Karoly, K.~Dell, K.~Leyde, M.~J. Cook, and M.~I. Maturana,
  ``Spatiotemporal {Patterns} of {High}-{Frequency} {Activity} (80-170 {Hz}) in
  {Long}-term {Intracranial} {EEG},'' {\em Neurology}, Dec. 2020.

\bibitem{proix_forecasting_2021}
T.~Proix, W.~Truccolo, M.~G. Leguia, T.~K. Tcheng, D.~King-Stephens, V.~R. Rao,
  and M.~O. Baud, ``Forecasting seizure risk in adults with focal epilepsy: a
  development and validation study,'' {\em The Lancet. Neurology}, vol.~20,
  pp.~127--135, Feb. 2021.

\bibitem{karoly_circadian_2017}
P.~J. Karoly, H.~Ung, D.~B. Grayden, L.~Kuhlmann, K.~Leyde, M.~J. Cook, and
  D.~R. Freestone, ``The circadian profile of epilepsy improves seizure
  forecasting,'' {\em Brain}, vol.~140, pp.~2169--2182, Aug. 2017.

\bibitem{scott_viability_2021}
J.~M. Scott, S.~V. Gliske, L.~Kuhlmann, and W.~C. Stacey, ``Viability of
  {Preictal} {High}-{Frequency} {Oscillation} {Rates} as a {Biomarker} for
  {Seizure} {Prediction},'' {\em Frontiers in Human Neuroscience}, vol.~14,
  p.~612899, Jan. 2021.

\bibitem{burrello_laelaps:_2019}
A.~Burrello, L.~Cavigelli, K.~Schindler, L.~Benini, and A.~Rahimi, ``Laelaps:
  {An} {Energy}-{Efficient} {Seizure} {Detection} {Algorithm} from {Long}-term
  {Human} {iEEG} {Recordings} without {False} {Alarms},'' in {\em 2019
  {Design}, {Automation} \& {Test} in {Europe} {Conference} \& {Exhibition}
  ({DATE})}, pp.~752--757, Mar. 2019.

\bibitem{lee_learning_1999}
D.~D. Lee and H.~S. Seung, ``Learning the parts of objects by non-negative
  matrix factorization,'' {\em Nature}, vol.~401, pp.~788--791, Oct. 1999.

\bibitem{atif_improved_2019}
S.~M. Atif, S.~Qazi, and N.~Gillis, ``Improved {SVD}-based initialization for
  nonnegative matrix factorization using low-rank correction,'' {\em Pattern
  Recognition Letters}, vol.~122, pp.~53--59, May 2019.

\bibitem{huang_empirical_1998}
N.~E. Huang, Z.~Shen, S.~R. Long, M.~C. Wu, H.~H. Shih, Q.~Zheng, N.-C. Yen,
  C.~C. Tung, and H.~H. Liu, ``The empirical mode decomposition and the
  {Hilbert} spectrum for nonlinear and non-stationary time series analysis,''
  {\em Proceedings of the Royal Society of London. Series A: Mathematical,
  Physical and Engineering Sciences}, vol.~454, pp.~903--995, Mar. 1998.

\bibitem{huang_confidence_2003}
N.~E. Huang, M.-L.~C. Wu, S.~R. Long, S.~S.~P. Shen, W.~Qu, P.~Gloersen, and
  K.~L. Fan, ``A {Confidence} {Limit} for the {Empirical} {Mode}
  {Decomposition} and {Hilbert} {Spectral} {Analysis},'' {\em Proceedings:
  Mathematical, Physical and Engineering Sciences}, vol.~459, no.~2037,
  pp.~2317--2345, 2003.

\bibitem{kaplan_nonstationary_2005}
A.~Y. Kaplan, A.~A. Fingelkurts, A.~A. Fingelkurts, S.~V. Borisov, and B.~S.
  Darkhovsky, ``Nonstationary nature of the brain activity as revealed by
  {EEG}/{MEG}: {Methodological}, practical and conceptual challenges,'' {\em
  Signal Processing}, vol.~85, pp.~2190--2212, Nov. 2005.

\bibitem{fingelkurts_operational_2001}
A.~A. Fingelkurts and A.~A. Fingelkurts, ``Operational {Architectonics} of the
  {Human} {Brain} {Biopotential} {Field}: {Towards} {Solving} the
  {Mind}-{Brain} {Problem},'' {\em Brain and Mind}, vol.~2, pp.~261--296, Dec.
  2001.

\bibitem{rehman_multivariate_2010}
N.~Rehman and D.~P. Mandic, ``Multivariate empirical mode decomposition,'' {\em
  Proceedings of the Royal Society A: Mathematical, Physical and Engineering
  Sciences}, vol.~466, pp.~1291--1302, May 2010.

\bibitem{huang_hilbert-huang_2014}
N.~E. Huang, {\em Hilbert-{Huang} {Transform} and {Its} {Applications}}.
\newblock World Scientific, 2014.

\bibitem{lv_multivariate_2016}
Y.~Lv, R.~Yuan, and G.~Song, ``Multivariate empirical mode decomposition and
  its application to fault diagnosis of rolling bearing,'' {\em Mechanical
  Systems and Signal Processing}, vol.~81, pp.~219--234, Dec. 2016.

\bibitem{hurley_comparing_2009}
N.~Hurley and S.~Rickard, ``Comparing {Measures} of {Sparsity},'' {\em IEEE
  Transactions on Information Theory}, vol.~55, pp.~4723--4741, Oct. 2009.

\bibitem{sakoe_dynamic_1978}
H.~Sakoe and S.~Chiba, ``Dynamic programming algorithm optimization for spoken
  word recognition,'' {\em IEEE Transactions on Acoustics, Speech, and Signal
  Processing}, vol.~26, pp.~43--49, Feb. 1978.

\bibitem{tibshirani_regression_1996}
R.~Tibshirani, ``Regression {Shrinkage} and {Selection} via the {Lasso},'' {\em
  Journal of the Royal Statistical Society. Series B (Methodological)},
  vol.~58, no.~1, pp.~267--288, 1996.

\bibitem{parrino_cyclic_2014}
L.~Parrino, A.~Grassi, and G.~Milioli, ``Cyclic alternating pattern in
  polysomnography: what is it and what does it mean?,'' {\em Current Opinion in
  Pulmonary Medicine}, vol.~20, pp.~533--541, Nov. 2014.

\bibitem{wu_study_2004}
Z.~Wu and N.~E. Huang, ``A study of the characteristics of white noise using
  the empirical mode decomposition method,'' {\em Proceedings of the Royal
  Society of London. Series A: Mathematical, Physical and Engineering
  Sciences}, vol.~460, pp.~1597--1611, June 2004.

\bibitem{flandrin_empirical_2004}
P.~Flandrin, G.~Rilling, and P.~Goncalves, ``Empirical mode decomposition as a
  filter bank,'' {\em IEEE Signal Processing Letters}, vol.~11, pp.~112--114,
  Feb. 2004.

\bibitem{ur_rehman_filter_2011}
N.~ur~Rehman and D.~P. Mandic, ``Filter {Bank} {Property} of {Multivariate}
  {Empirical} {Mode} {Decomposition},'' {\em IEEE Transactions on Signal
  Processing}, vol.~59, pp.~2421--2426, May 2011.

\bibitem{shah_characterizing_2019}
P.~Shah, A.~Ashourvan, F.~Mikhail, A.~Pines, L.~Kini, K.~Oechsel, S.~R. Das,
  J.~M. Stein, R.~T. Shinohara, D.~S. Bassett, B.~Litt, and K.~A. Davis,
  ``Characterizing the role of the structural connectome in seizure dynamics,''
  {\em Brain}, vol.~142, pp.~1955--1972, July 2019.

\bibitem{scheich_interval_1969}
H.~Scheich, ``Interval histograms and periodic diurnal changes of human alpha
  rhythms,'' {\em Electroencephalography and Clinical Neurophysiology},
  vol.~26, p.~442, Apr. 1969.

\bibitem{spencer_circadian_2016}
D.~C. Spencer, F.~T. Sun, S.~N. Brown, B.~C. Jobst, N.~B. Fountain, V.~S.~S.
  Wong, E.~A. Mirro, and M.~Quigg, ``Circadian and ultradian patterns of
  epileptiform discharges differ by seizure-onset location during long-term
  ambulatory intracranial monitoring,'' {\em Epilepsia}, vol.~57,
  pp.~1495--1502, Sept. 2016.

\bibitem{smyk_circadian_2020}
M.~K. Smyk and G.~van Luijtelaar, ``Circadian {Rhythms} and {Epilepsy}: {A}
  {Suitable} {Case} for {Absence} {Epilepsy},'' {\em Frontiers in Neurology},
  vol.~11, 2020.

\bibitem{cummings_diurnal_2000}
L.~Cummings, A.~Dane, J.~Rhodes, P.~Lynch, and A.~M. Hughes, ``Diurnal
  variation in the quantitative {EEG} in healthy adult volunteers,'' {\em
  British Journal of Clinical Pharmacology}, vol.~50, pp.~21--26, July 2000.

\bibitem{kaiser_ultradian_2008}
D.~A. Kaiser, ``Ultradian and {Circadian} {Effects} in {Electroencephalography}
  {Activity},'' {\em Biofeedback}, vol.~36, no.~4, p.~148, 2008.

\bibitem{chapotot_high_2000}
F.~Chapotot, C.~Jouny, A.~Muzet, A.~Buguet, and G.~Brandenberger, ``High
  frequency waking {EEG}: reflection of a slow ultradian rhythm in daytime
  arousal,'' {\em NeuroReport}, vol.~11, pp.~2223--2227, July 2000.

\bibitem{maturana_critical_2020}
M.~I. Maturana, C.~Meisel, K.~Dell, P.~J. Karoly, W.~D’Souza, D.~B. Grayden,
  A.~N. Burkitt, P.~Jiruska, J.~Kudlacek, J.~Hlinka, M.~J. Cook, L.~Kuhlmann,
  and D.~R. Freestone, ``Critical slowing down as a biomarker for seizure
  susceptibility,'' {\em Nature Communications}, vol.~11, May 2020.

\bibitem{bernard_circadianmultidien_2021}
C.~Bernard, ``Circadian/multidien {Molecular} {Oscillations} and {Rhythmicity}
  of {Epilepsy} ({MORE}),'' {\em Epilepsia}, vol.~62, no.~S1, pp.~S49--S68,
  2021.

\bibitem{karoly_cycles_2021}
P.~J. Karoly, V.~R. Rao, N.~M. Gregg, G.~A. Worrell, C.~Bernard, M.~J. Cook,
  and M.~O. Baud, ``Cycles in epilepsy,'' {\em Nature Reviews Neurology},
  vol.~17, pp.~267--284, May 2021.

\bibitem{rao_cues_2021}
V.~R. Rao, M.~G. Leguia, T.~K. Tcheng, and M.~O. Baud, ``Cues for seizure
  timing,'' {\em Epilepsia}, vol.~62, no.~S1, pp.~S15--S31, 2021.

\bibitem{badawy_epilepsy_2012}
R.~A.~B. Badawy, D.~R. Freestone, A.~Lai, and M.~J. Cook, ``Epilepsy:
  {Ever}-changing states of cortical excitability,'' {\em Neuroscience},
  vol.~222, pp.~89--99, Oct. 2012.

\bibitem{meisel_intrinsic_2015}
C.~Meisel, A.~Schulze-Bonhage, D.~Freestone, M.~J. Cook, P.~Achermann, and
  D.~Plenz, ``Intrinsic excitability measures track antiepileptic drug action
  and uncover increasing/decreasing excitability over the wake/sleep cycle,''
  {\em Proceedings of the National Academy of Sciences}, vol.~112,
  pp.~14694--14699, Nov. 2015.

\bibitem{battista_application_2007}
B.~M. Battista, C.~Knapp, T.~McGee, and V.~Goebel, ``Application of the
  empirical mode decomposition and {Hilbert}-{Huang} transform to seismic
  reflection data,'' {\em GEOPHYSICS}, vol.~72, pp.~H29--H37, Mar. 2007.

\bibitem{hu_soil_2013}
W.~Hu and B.~C. Si, ``Soil water prediction based on its scale-specific control
  using multivariate empirical mode decomposition,'' {\em Geoderma},
  vol.~193-194, pp.~180--188, Feb. 2013.

\bibitem{huang_application_2013}
J.-R. Huang, S.-Z. Fan, M.~F. Abbod, K.-K. Jen, J.-F. Wu, and J.-S. Shieh,
  ``Application of {Multivariate} {Empirical} {Mode} {Decomposition} and
  {Sample} {Entropy} in {EEG} {Signals} via {Artificial} {Neural} {Networks}
  for {Interpreting} {Depth} of {Anesthesia},'' {\em Entropy}, vol.~15,
  pp.~3325--3339, Sept. 2013.

\bibitem{rojas_application_2013}
A.~Rojas, J.~M. Górriz, J.~Ramírez, I.~A. Illán, F.~J. Martínez-Murcia,
  A.~Ortiz, M.~Gómez~Río, and M.~Moreno-Caballero, ``Application of
  {Empirical} {Mode} {Decomposition} ({EMD}) on {DaTSCAN} {SPECT} images to
  explore {Parkinson} {Disease},'' {\em Expert Systems with Applications},
  vol.~40, pp.~2756--2766, June 2013.

\bibitem{xue_adaptively_2015}
X.~Xue, J.~Zhou, Y.~Xu, W.~Zhu, and C.~Li, ``An adaptively fast ensemble
  empirical mode decomposition method and its applications to rolling element
  bearing fault diagnosis,'' {\em Mechanical Systems and Signal Processing},
  vol.~62-63, pp.~444--459, Oct. 2015.

\bibitem{deering_use_2005}
R.~Deering and J.~Kaiser, ``The use of a masking signal to improve empirical
  mode decomposition,'' in {\em Proceedings. ({ICASSP} '05). {IEEE}
  {International} {Conference} on {Acoustics}, {Speech}, and {Signal}
  {Processing}, 2005.}, vol.~4, pp.~iv/485--iv/488 Vol. 4, Mar. 2005.
\newblock ISSN: 2379-190X.

\bibitem{li_improved_2015}
H.~Li, C.~Wang, and D.~Zhao, ``An {Improved} {EMD} and {Its} {Applications} to
  {Find} the {Basis} {Functions} of {EMI} {Signals},'' {\em Mathematical
  Problems in Engineering}, vol.~2015, p.~e150127, Nov. 2015.

\bibitem{stam_nonlinear_2005}
C.~J. Stam, ``Nonlinear dynamical analysis of {EEG} and {MEG}: {Review} of an
  emerging field,'' {\em Clinical Neurophysiology}, vol.~116, pp.~2266--2301,
  Oct. 2005.

\bibitem{vidaurre_brain_2017}
D.~Vidaurre, S.~M. Smith, and M.~W. Woolrich, ``Brain network dynamics are
  hierarchically organized in time,'' {\em Proceedings of the National Academy
  of Sciences}, vol.~114, pp.~12827--12832, Nov. 2017.

\bibitem{badawy_peri-ictal_2009}
R.~Badawy, R.~Macdonell, G.~Jackson, and S.~Berkovic, ``The peri-ictal state:
  cortical excitability changes within 24 h of a seizure,'' {\em Brain: A
  Journal of Neurology}, vol.~132, pp.~1013--1021, Apr. 2009.

\bibitem{arzy_antiepileptic_2010}
S.~Arzy, G.~Allali, D.~Brunet, C.~M. Michel, P.~W. Kaplan, and M.~Seeck,
  ``Antiepileptic drugs modify power of high {EEG} frequencies and their neural
  generators,'' {\em European Journal of Neurology}, vol.~17, pp.~1308--1312,
  Oct. 2010.

\bibitem{carney_chronotherapy_2014}
P.~Carney, D.~Stanley, and S.~Talathi, ``Chronotherapy in the treatment of
  epilepsy,'' {\em ChronoPhysiology and Therapy}, p.~109, Nov. 2014.

\bibitem{manganaro_need_2017}
S.~Manganaro, T.~Loddenkemper, and A.~Rotenberg, ``The {Need} for
  {Antiepileptic} {Drug} {Chronotherapy} to {Treat} {Selected} {Childhood}
  {Epilepsy} {Syndromes} and {Avert} the {Harmful} {Consequences} of {Drug}
  {Resistance},'' {\em Journal of Central Nervous System Disease}, vol.~9, Dec.
  2017.

\bibitem{ramgopal_chronopharmacology_2013}
S.~Ramgopal, S.~Thome-Souza, and T.~Loddenkemper, ``Chronopharmacology of
  {Anti}-{Convulsive} {Therapy},'' {\em Current Neurology and Neuroscience
  Reports}, vol.~13, p.~339, Mar. 2013.

\bibitem{leguia_seizure_2021}
M.~G. Leguia, R.~G. Andrzejak, C.~Rummel, J.~M. Fan, E.~A. Mirro, T.~K. Tcheng,
  V.~R. Rao, and M.~O. Baud, ``Seizure {Cycles} in {Focal} {Epilepsy},'' {\em
  JAMA neurology}, vol.~78, pp.~454--463, Apr. 2021.

\end{thebibliography}
\bibliographystyle{ieeetr}

\newpage
\section*{Supporting Information}
\renewcommand{\thefigure}{S\arabic{figure}}
\setcounter{figure}{0}
\counterwithin{figure}{subsection}
\counterwithin{table}{subsection}
\renewcommand\thesection{Supporting S\arabic{section}}
\setcounter{section}{0}

\section{Subject data information \label{suppl:tableData}}

\begin{table}[ht!]
\centering
\begin{tabular}{lcccc}
\textbf{Subject} & \multicolumn{1}{l}{\textbf{Duration}} & \multicolumn{1}{l}{\textbf{Duration}} & \multicolumn{1}{l}{\textbf{Number of}} & \multicolumn{1}{l}{\textbf{Mean seizure}}  \\
 & \multicolumn{1}{l}{\textbf{in hours}} & \multicolumn{1}{l}{\textbf{in days}} & \multicolumn{1}{l}{\textbf{seizures}} & \multicolumn{1}{l}{\textbf{duration (minutes)}}  \\
\hline
\textbf{ID01}    & 293                                     & 12                                     & 2                               & 10.030                                         \\
\textbf{ID02}    & 235                                     & 10                                     & 2                               & 1.468                                          \\
\textbf{ID03}    & 158                                     & 7                                      & 4                               & 1.078                                          \\
\textbf{ID04}    & 41                                      & 2                                      & 14                              & 0.699                                          \\
\textbf{ID05}    & 109                                     & 5                                      & 4                               & 0.278                                          \\
\textbf{ID06}    & 146                                     & 6                                      & 8                               & 0.765                                          \\
\textbf{ID07}    & 69                                      & 3                                      & 4                               & 1.159                                          \\
\textbf{ID08}    & 144                                     & 6                                      & 70                              & 0.366                                          \\
\textbf{ID09}    & 41                                      & 2                                      & 27                              & 0.706                                          \\
\textbf{ID10}    & 42                                      & 2                                      & 17                              & 1.181                                          \\
\textbf{ID11}    & 212                                     & 9                                      & 2                               & 1.526                                          \\
\textbf{ID12}    & 191                                     & 8                                      & 9                               & 2.441                                          \\
\textbf{ID13}    & 104                                     & 4                                      & 7                               & 1.717                                          \\
\textbf{ID14}    & 161                                     & 7                                      & 60                              & 0.430                                          \\
\textbf{ID15}    & 196                                     & 8                                      & 2                               & 1.576                                          \\
\textbf{ID16}    & 177                                     & 7                                      & 5                               & 3.174                                          \\
\textbf{ID17}    & 130                                     & 5                                      & 2                               & 1.632                                          \\
\textbf{ID18}    & 205                                     & 9                                      & 5                               & 3.319                                          \\
\hline
\textbf{Total}   & 2656                                    & 111                                    & 244                             &     \\
\hline
\textbf{Average}   &                                     &                                     &                              &    1.864
\end{tabular}
\caption{\label{tab:descr_data}For each subject the following information is provided: \textbf{Duration in hours:} duration of iEEG recordings in hours. \textbf{Duration in days:} duration of iEEG recordings in days. \textbf{Number of seizures:} number of subject's seizures annotated. \textbf{Mean seizure duration:} mean seizure duration across all annotated seizures in minutes.}
\end{table}

\newpage
\section{Visualising Seizure Dissimilarity}\label{suppl:visualDissMat}

\begin{figure}[h!]
    \hspace{-1cm}
    \includegraphics[scale = 1]{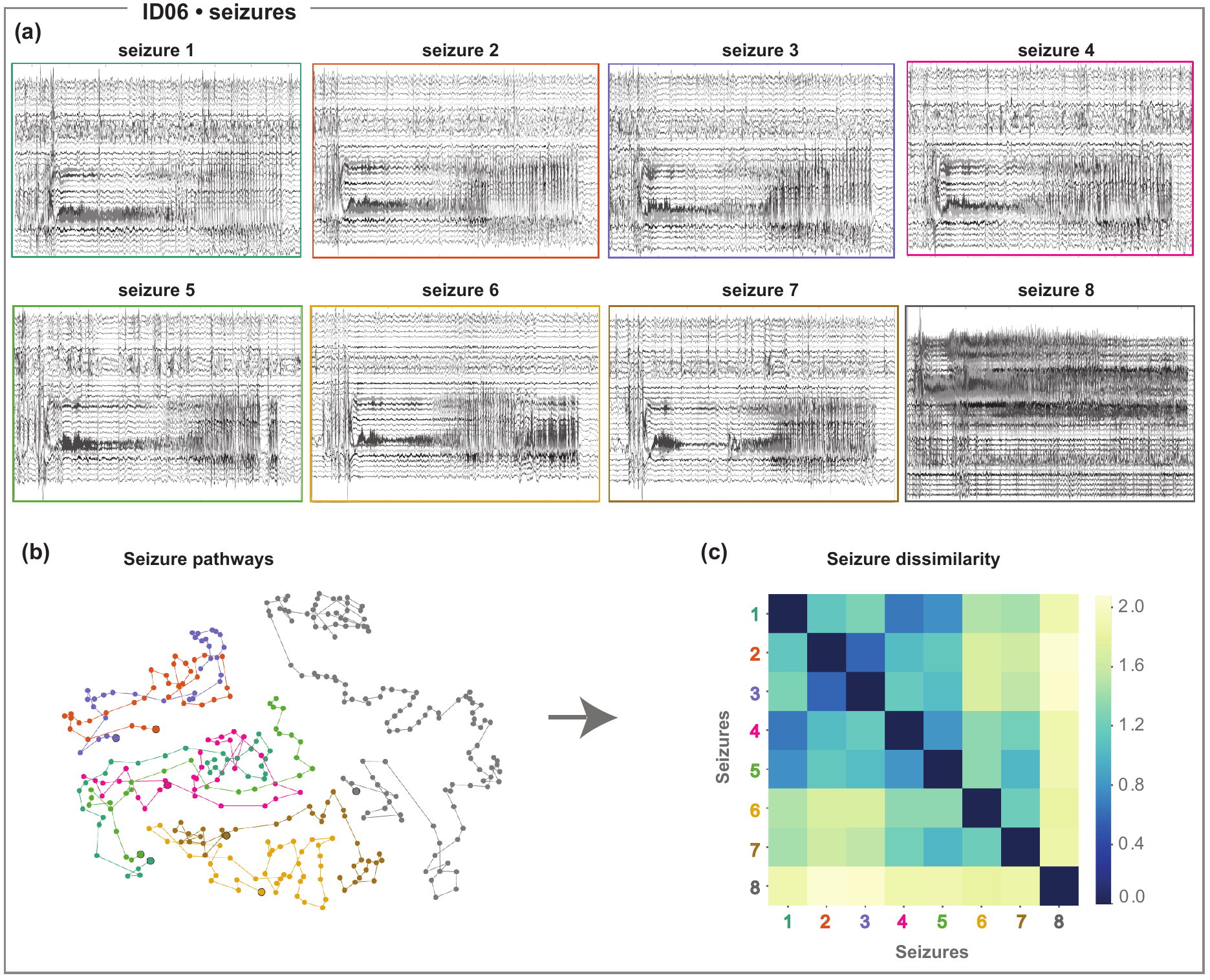}
     \caption{\textbf{Seizure dissimilarity matrix for example subject ID06.} (a) Seizure iEEG traces are shown in the top panel of the figure. (b) Functional network evolution of all seizures projected into 2D space using multi-dimensional scaling (MDS) for visualisation of seizure pathways (see \cite{schroeder_seizure_2020} for details). Similar seizures tend to be placed closer together in this projection. Seizures are displayed with distinct colours to distinguish seizure events. The starting points of seizures are marked with a black outline circle. (c) Seizure dissimilarity matrix, capturing the differences in seizure pathways between each pair of seizures.}
     \label{fig:complfigure}
\end{figure}

The iEEG traces of all seizures for subject ID06 are shown in Fig.~\ref{fig:complfigure}a for visual comparison of the different seizures and the quantified seizure evolution differences displayed as trajectories (Fig.~\ref{fig:complfigure}b) along with the dissimilarity values (Fig.~\ref{fig:complfigure}c) for each pair of seizures. The bottom panel of the figure is similar to Fig.~\ref{fig:mantel}a,b.  

\newpage

\section{Association between Seizure Dissimilarity and Seizure Band Power Distance}\label{suppl:Corr_band_power}

Here, we want to demonstrate that the band power signal itself does not explain how seizure pathways change, or at least not as well as specific fluctuations of bandpower on particular timescales (as we presented in the main results). To this end, we associated each subject’s pathway dissimilarity matrix with differences in the raw band power signal (termed band power distance from now on). 

In order to explore if band power distance explains seizure dissimilarity, we applied a linear regression framework. We implemented two models: (i) using the seizure onset time window and (ii) using the time window just before the seizure onset (onset window - 1). 

For consistency with the main part of our analysis (see Fig.~\ref{fig:mantel}a\&b) and allow comparison with findings in Fig.~\ref{fig:lasso}, a pairwise band power distance was obtained as the euclidean distance from the data matrix $X$ shown in Fig.~\ref{fig:NMF_illustration}b. In other words, the band power distance is the euclidean distance between two time windows in terms of their band power in all frequency bands and channels.

\begin{figure}[h!]
    \hspace{-1cm}
    \includegraphics[scale = 1]{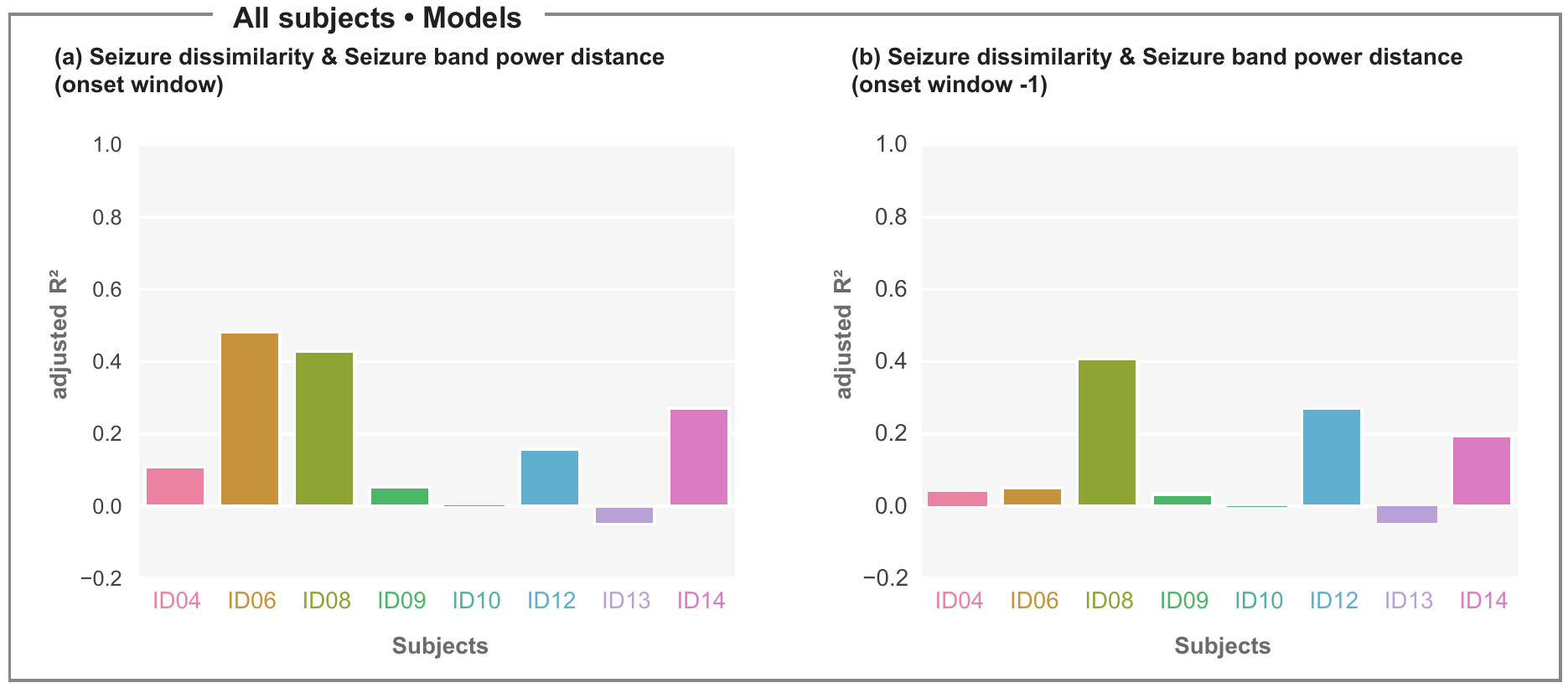}
       \caption{\textbf{Relating seizure dissimilarity with seizure band power distance.} (a)\&(b) Summary across subjects represented with bar charts of the adjusted $R^2$ values obtained from the linear models using the onset window (Left plot: (a)) and the onset window -1 (Right plot: (b)).}
    \label{fig:band_power_corr}
\end{figure}

As can be seen in Fig.~\ref{fig:band_power_corr}, the low values of adjusted $R^2$ across all subjects indicate the band power signal itself does not explain how seizures change over time. Instead the decomposition of the band power signal into fluctuations of particular timescales is crucial, and only some of these timescales contribute explanatory power, as we have shown in Fig.~\ref{fig:lasso}.

\newpage

\section{Tests for statistical significance in model $R^{2}$\label{suppl:signtest}}
\subsection{Randomly shifted onset times}

\begin{figure}[h!]
    \hspace{-1cm}
    \includegraphics[scale = 1]{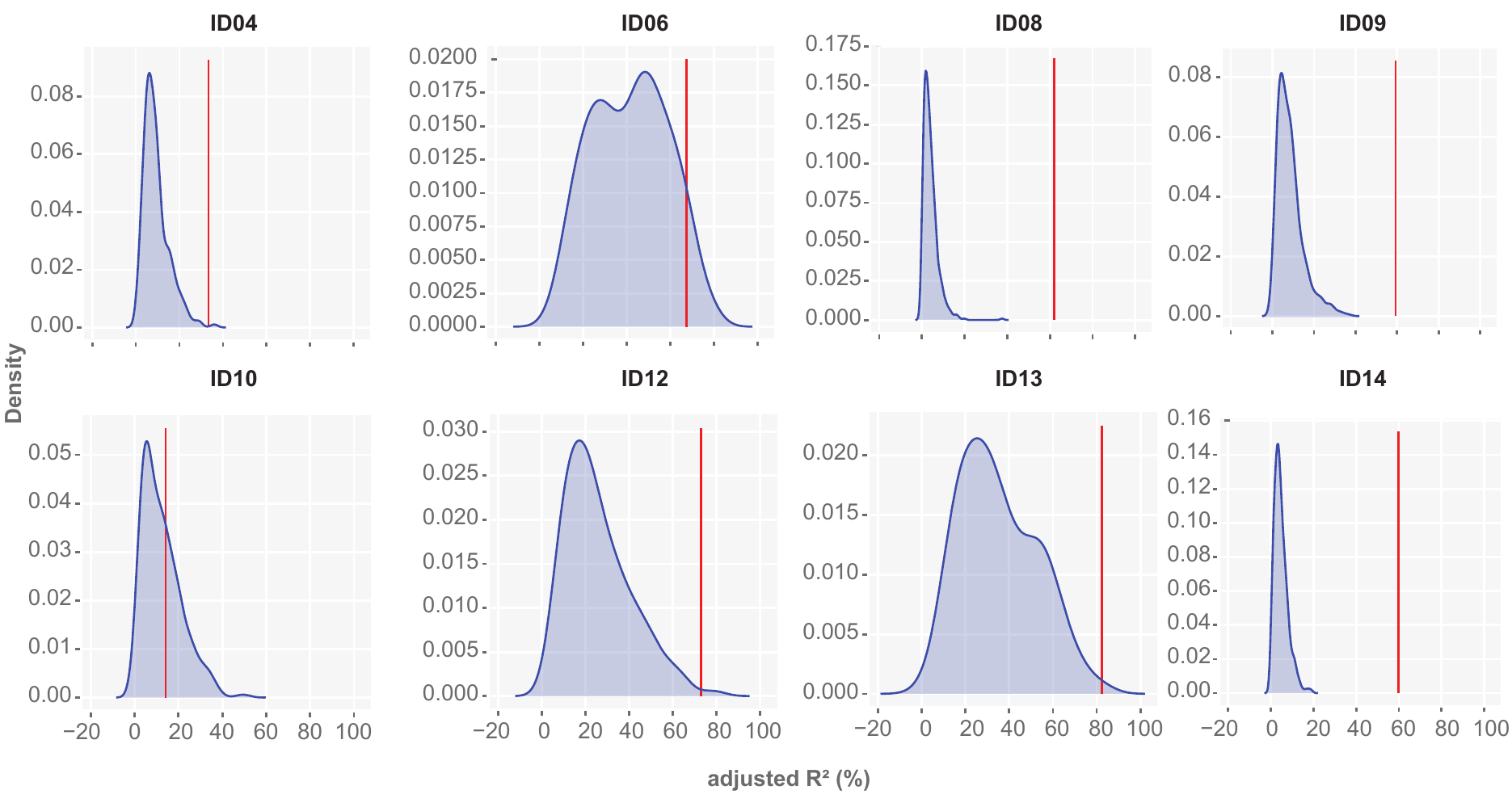}
       \caption{Distribution of the adjusted $R^2$ values using random seizure timings, after implementing the positive LASSO and OLS regression model for each subject. The seizure dissimilarity matrix was used as response variable, while the seizure IMF distance matrices for the random seizure times and the seizure temporal distance were included in the model as explanatory variables. The vertical red line represents the adjusted $R^{2}$ for the same analysis performed on the original seizure onset times (see Section~\ref{subsec:Linear analysis} and Fig.~\ref{fig:lasso}d).}
    \label{fig:adjusted_Rsquared}
\end{figure}

We randomly shifted seizure onset times to test if the multiple linear regression model $R^2$ values would have occurred by chance. Adjusted $R^2$ values for 500 iterations, along with the actual adjusted $R^2$ are shown in Fig.~\ref{fig:adjusted_Rsquared}.

\subsection{Randomly permuted onset times\label{subsec:permszorder}}

\begin{figure}[h!]
    \hspace{-1cm}
    \includegraphics[scale = 1]{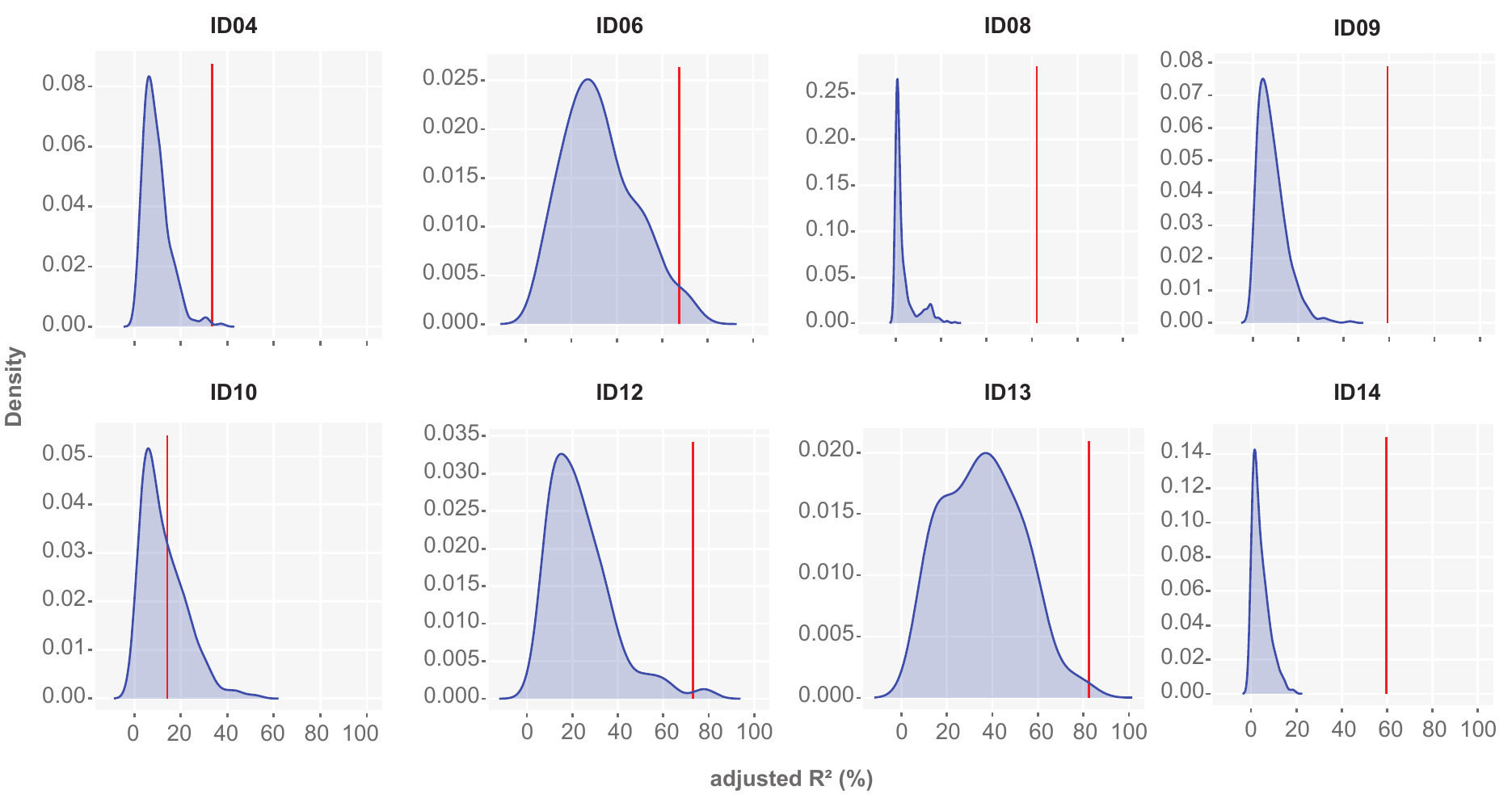}
       \caption{Distribution of the adjusted $R^2$ values using permuted seizure orders. The seizure dissimilarity matrix was used as response variable, while the seizure IMF distance matrices for the random seizures and the seizure temporal distance were included in the model as explanatory variables. The vertical red line represents the adjusted $R^2$ for the same analysis performed on the original seizure order (see Section~\ref{subsec:Linear analysis} and Fig.~\ref{fig:lasso}d).}
    \label{fig:adjusted_Rsquared_shuffle}
\end{figure}

Similarly to the analysis described in the previous section, we additionally performed a permutation test. In each permutation iteration, we first randomly permuted the order of the seizures (but not their timing). Then, we obtained new IMF seizure distance matrices and performed the LASSO and linear regression, exactly as described in the Section~\ref{subsec:Linear analysis}, leaving the response variable unchanged. Finally, we calculated the adjusted $R^2$ for each iteration (see Fig.~\ref{fig:adjusted_Rsquared_shuffle}). Statistical significance was determined based on a significance level of $5\%$. Again, the aforementioned steps were performed for all subjects with at least six recorded seizures. Significance levels were similarly for all tested subjects as in the previous section.

\section{Gini index for frequency bands $\theta, \alpha, \beta$ and $\gamma$}

\begin{figure}[h!]
    \hspace{-0.8cm}
    \includegraphics[scale = 1]{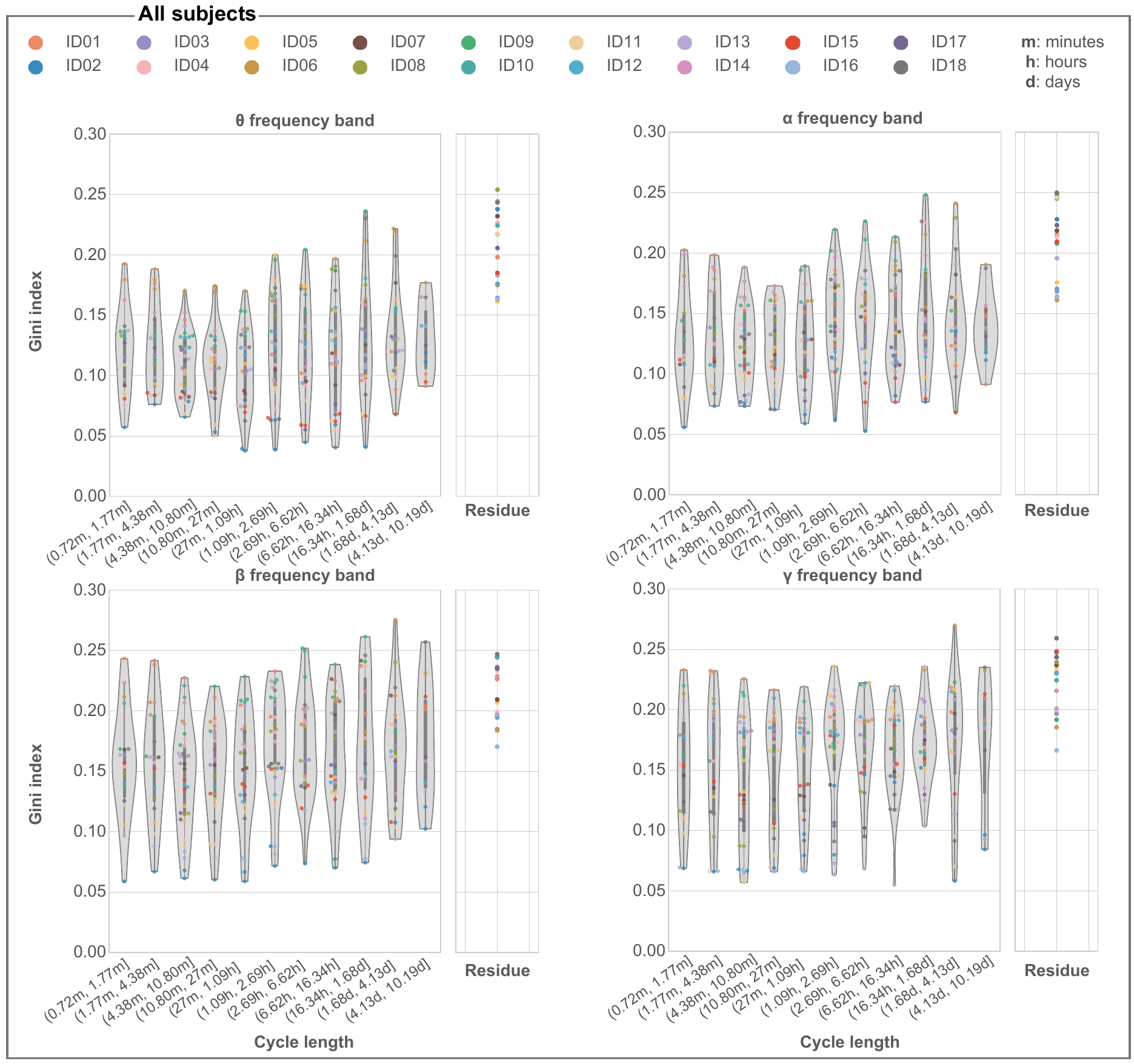}
       \caption{\textbf{Supporting Gini Index results  for the $\theta, \alpha, \beta$ and $\gamma$ frequency bands across all subjects.} Equivalent figure to Fig.~\ref{fig:Gini_index}}
    \label{fig:Gini_index_rest}
\end{figure}

As described in Section~\ref{subsec:spatial_heterogeneity}, we obtained the Gini index for the frequency bands: $\theta, \alpha, \beta$ and $\gamma$ (see Fig.~\ref{fig:Gini_index_rest}) in the same way as for $\delta$ band (Fig.~\ref{fig:Gini_index}).

\clearpage

\section{Choice of tuning parameter $\lambda$ for LASSO}

\begin{figure}[h]
    \centering
    \includegraphics[scale = 1]{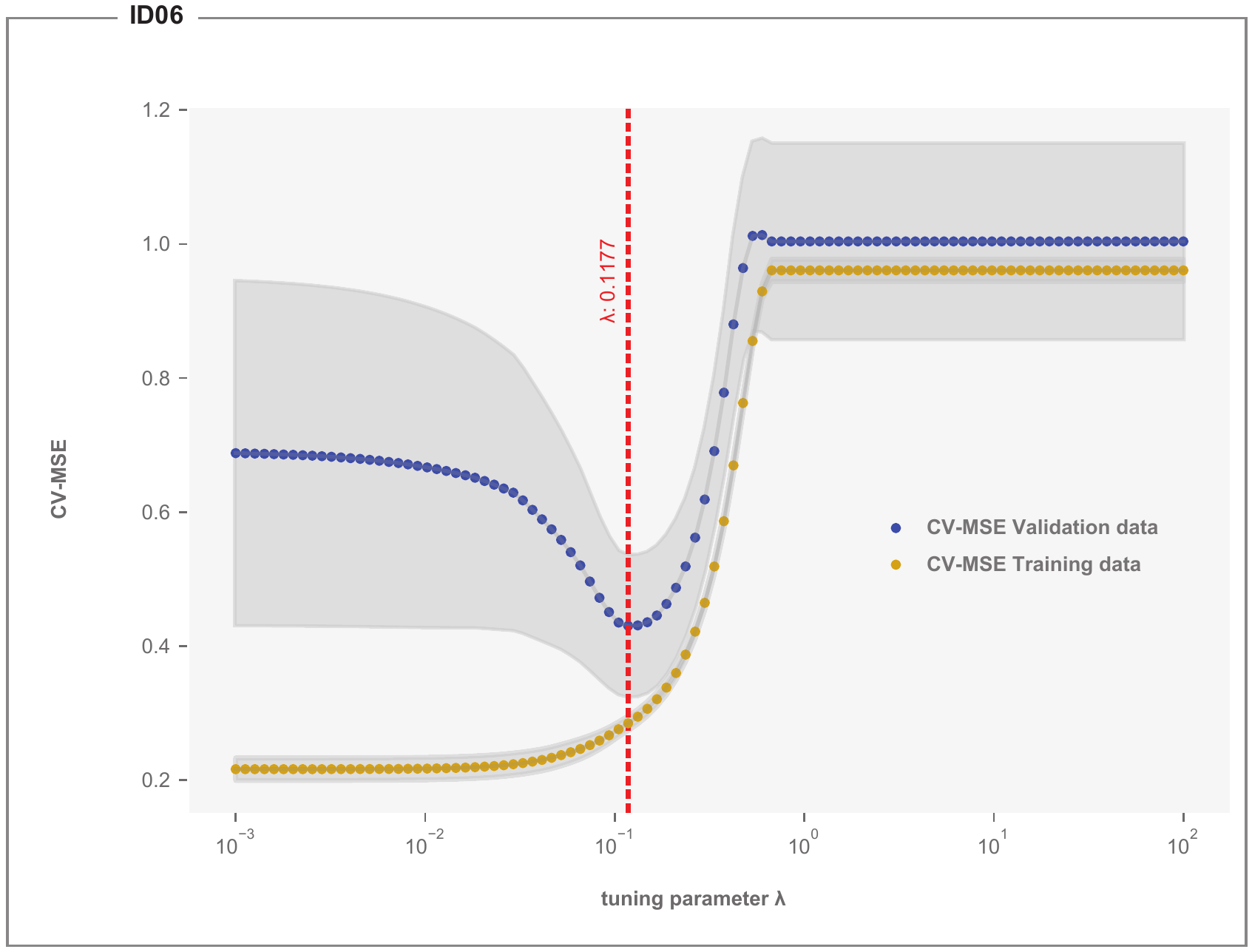}
    \caption{Cross-validation MSEs for the application of positive LASSO regression for example subject ID06. For each value of the tuning parameter $\lambda$, the CV-MSEs across the 10 folds are displayed in blue along with error bars which cover the mean plus or minus one standard error. Training MSE is displayed in yellow. The red vertical line represents the selected $\lambda$ value that corresponds to the minimum Cross-Validation MSE for the validated data set.}
    \label{fig:cv_error}
\end{figure}

\begin{figure}
    \centering
    \includegraphics[scale = 1]{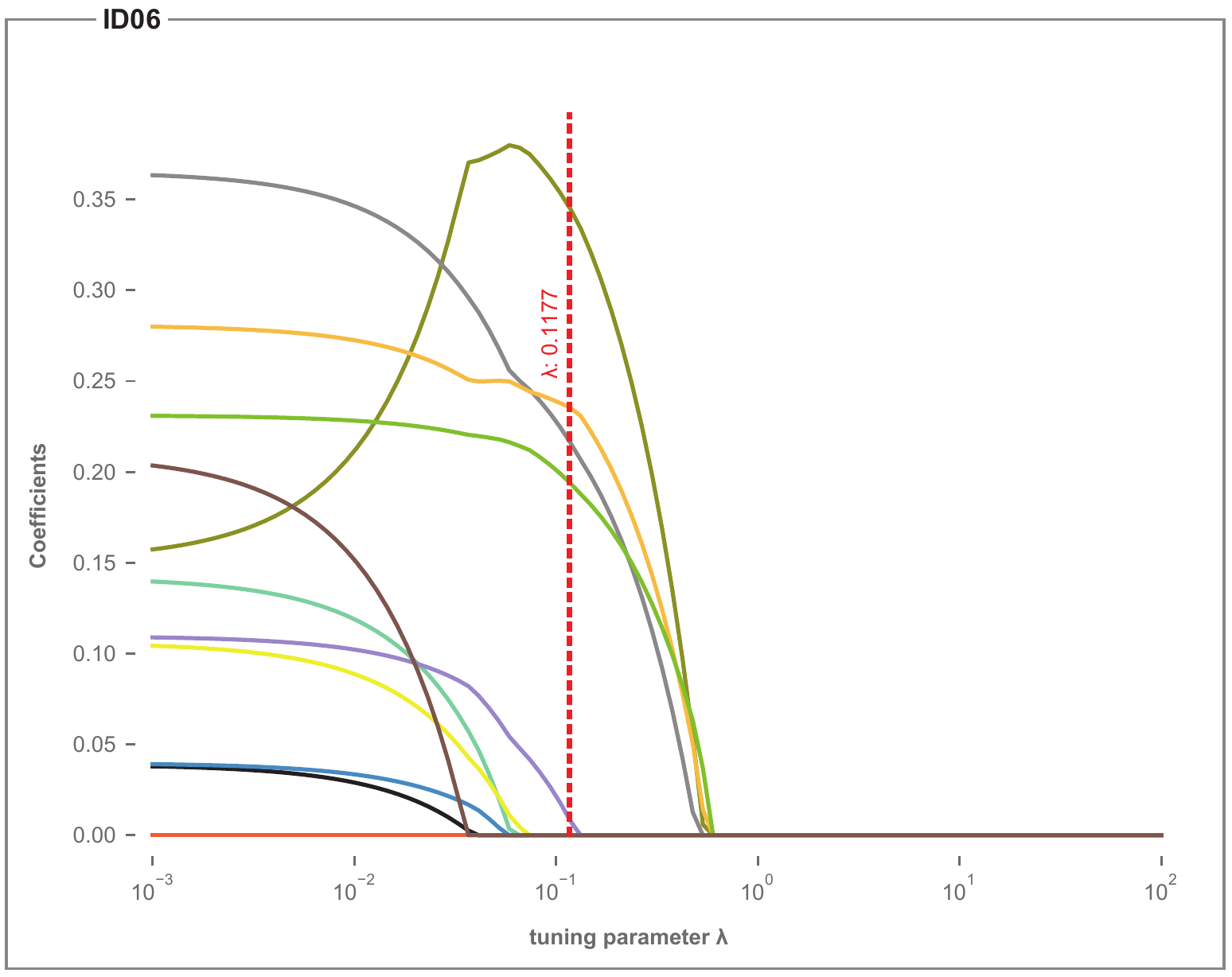}
    \caption{The effect of the tuning parameter $\lambda$ on positive LASSO regression coefficients for subject ID06. Each line represents the regression coefficient estimate for each explanatory variable. The red vertical line corresponds to the $\lambda$ parameter selected based on 10-fold cross validation approach.}
    \label{fig:tuning_parameters}
\end{figure}

Figures~\ref{fig:cv_error} \& \ref{fig:tuning_parameters} are complementary plots supporting the intermediate steps of the analysis described in Section~\ref{subsec:Linear analysis}. We implemented a 10-fold cross validation for choosing the best $\lambda$ parameter in LASSO. We chose a $\lambda$ that minimize the Cross-Validation Mean Square Error (CV-MSE) for the validated data set. The CV-MSE error for the training data set is also presented in Fig.~\ref{fig:cv_error} for reference for one example subject, ID06. 

\clearpage
\section{Determining which IMF fluctuations overlap with noise\label{subsec:noisyIMF}}

In order to evaluate if each of the IMFs is a good representation of fluctuations present in the data, we implemented an empirical analysis based on permutation resampling of the original time series ($H$ expression coefficients). We used this nullmodel to produce distributions of IMF fluctuation frequencies that would be expected from noise. In our nullmodel, we assumed a permuted time series (permuting time points in $H$) to represent noise. I.e. we preserve the distribution of the values in the time series, but the temporal fluctuations are destroyed through the permutation.

We randomly shuffled the columns (timing) of the $H$ matrix over 50 iterations and performed the MEMD analysis for each iteration. Then, for each IMF, in each iteration, we estimated the 2D distribution of the instantaneous frequency, and instantaneous amplitude (across all time points). We formed the average distribution across all iterations. We repeated the same analysis for the original (non-shuffled) data, for each IMF. Therefore we can calculate overlap in the distributions (between original and shuffled data). 

We used a 2D grid of (frequency, amplitude) with 800 frequency bin between $10^{-3}$ to $10^{4}$ in logarithmic scale, and 400 amplitude bins between $10^{-4}$ to $10^{0}$ in logarithmic scale (logarithmic scales of base 10 were used). In each (frequency, amplitude) bin, where the original signal overlapped with the permuted signal, the corresponding data points were labelled as overlapping with noise. These points can subsequently be removed from the calculation of the marginal Hilbert-Huang spectra for the original signal. These newly obtained marginal Hilbert-Huang spectra, excluding data points overlapping with noise, are shown in red in (Fig.~\ref{fig:PSD1}~\&~\ref{fig:PSD2}).

\begin{figure}[h!]
    \hspace{-0.8cm}
    \includegraphics[scale = 1]{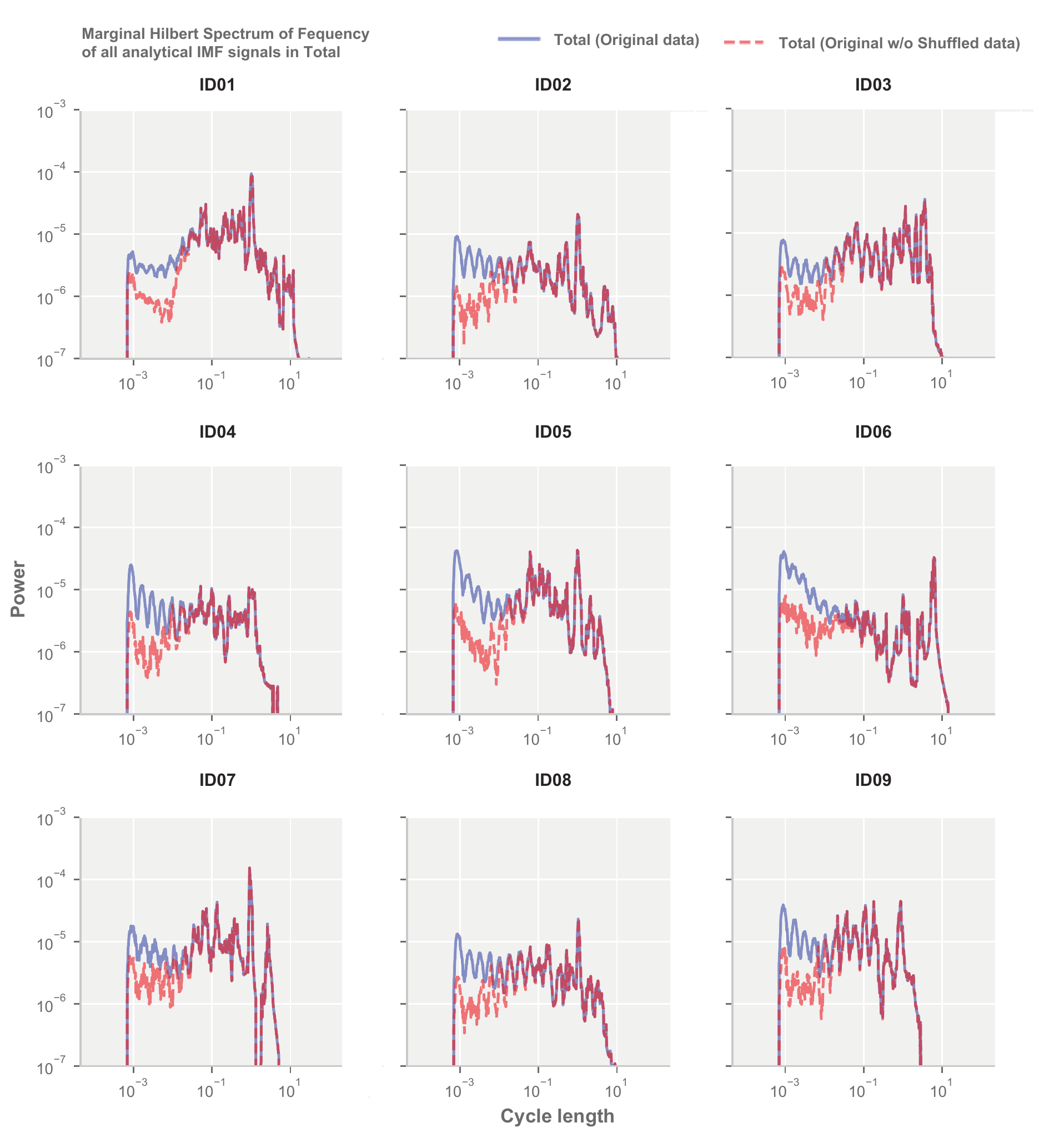}
       \caption{Marginal Hilbert spectrum of frequency for all analytical IMF signals across all dimensions for both the original data (blue line) and after excluding frequency-amplitude data points overlapping with noise (red line). Each panel corresponds to one subject.}
    \label{fig:PSD1}
\end{figure}

\begin{figure}[h!]
    \hspace{-0.8cm}
    \includegraphics[scale = 1]{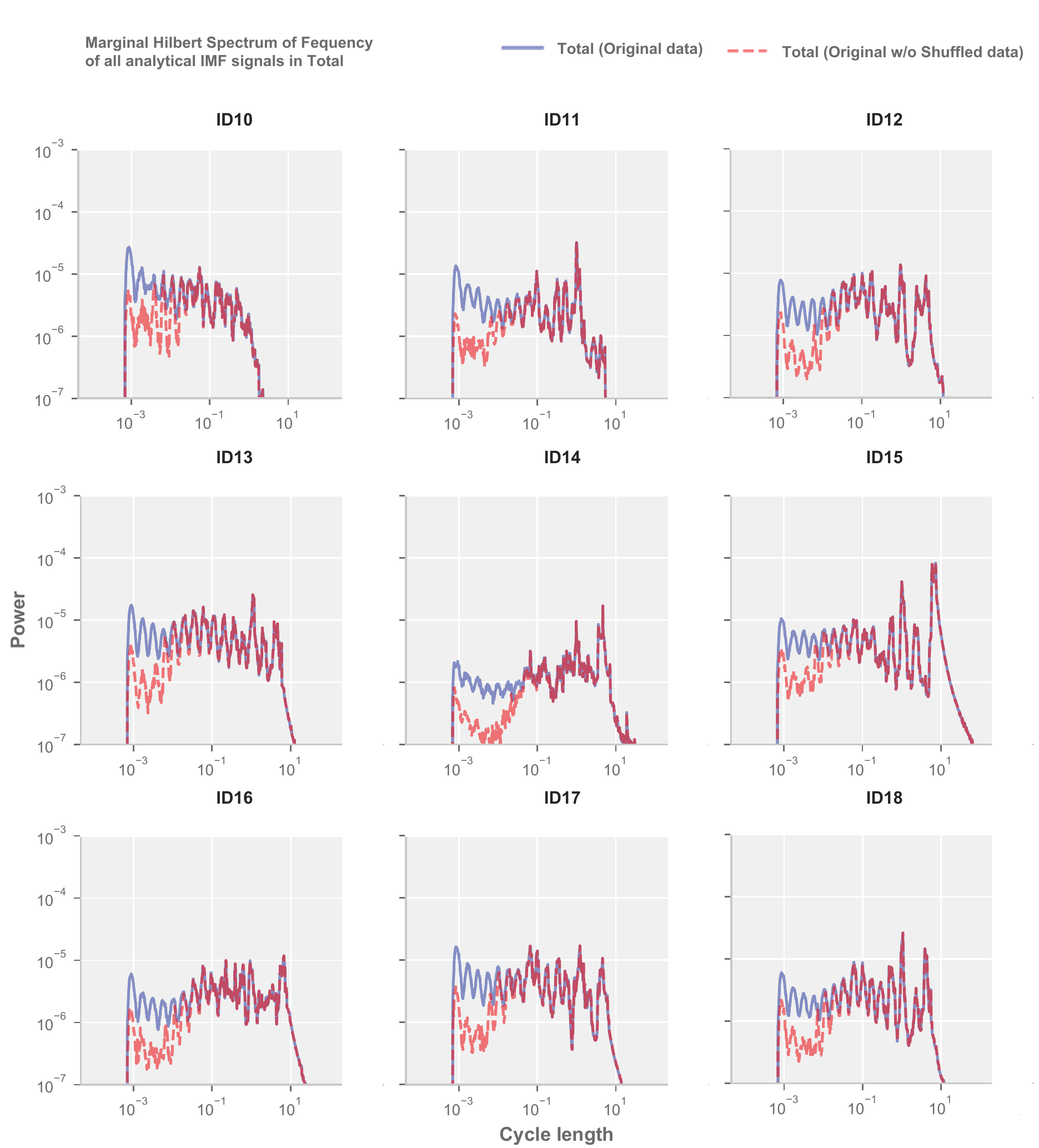}
       \caption{Continued: Marginal Hilbert spectrum of frequency for all analytical IMF signals across all dimensions for both the original data (blue line) and after excluding frequency-amplitude data points overlapping with noise (red line). Each panel corresponds to one subject.}
    \label{fig:PSD2}
\end{figure}

In Fig.~\ref{fig:PSD1}~\&~\ref{fig:PSD2}, we can clearly see that many frequencies of fluctuations in IMF1-3 are overlapping with noise in most subjects. The slower IMFs do not appear to be affected (as noise-IMF tend to decrease in amplitude for slower IMFs). While we could discard faster IMFs as noise due to the overlap, it is worth noting that these faster IMFs could carry some true fluctuation that is simply on the same timescale and of the same amplitude as the noise. This would be impossible to distinguish here, and therefore we present all results on all IMFs in the main text and will present supporting results with the faster IMFs removed in \ref{subsec:Raw_analysis-noise}.



\clearpage
\section{Alternative models for explaining the diversity in within-subject seizure evolutions}

We tested additional models to see how well they explain seizure variability, using the same framework as is described in Section~\ref{subsec:Linear analysis}.

\subsection{Association between seizure dissimilarity \& IMF seizure distance based on the time window before the seizure\label{subsec:raw_analysis_onset-1}}

To further validate our model in terms of the time window chosen for obtaining the seizure IMF distances, we performed an additional analysis using one time window before the window containing the seizure onset (termed onset window-1). The reasoning is that the IMF distances obtained in this manner cannot be containing any seizure-related changes in band power. 
As can be seen in Fig.~\ref{fig:R_squared_all_models}a and in the first two columns of Table~\ref{tab:R^2_Raw_Analysis}, the adjusted $R^2$, as well as the coefficient estimates and the IMF components remaining in the model for each subject are in agreement with the model shown in Fig.~\ref{fig:lasso}d. Thus, both models perform similarly indicating that the IMF distance results are robust towards changing a single window.

\subsection{Association between seizure dissimilarity \& IMF seizure distance excluding noise\label{subsec:Raw_analysis-noise}}

\begin{figure}[h!]
    \hspace{-0.75cm}
    \includegraphics[scale = 1]{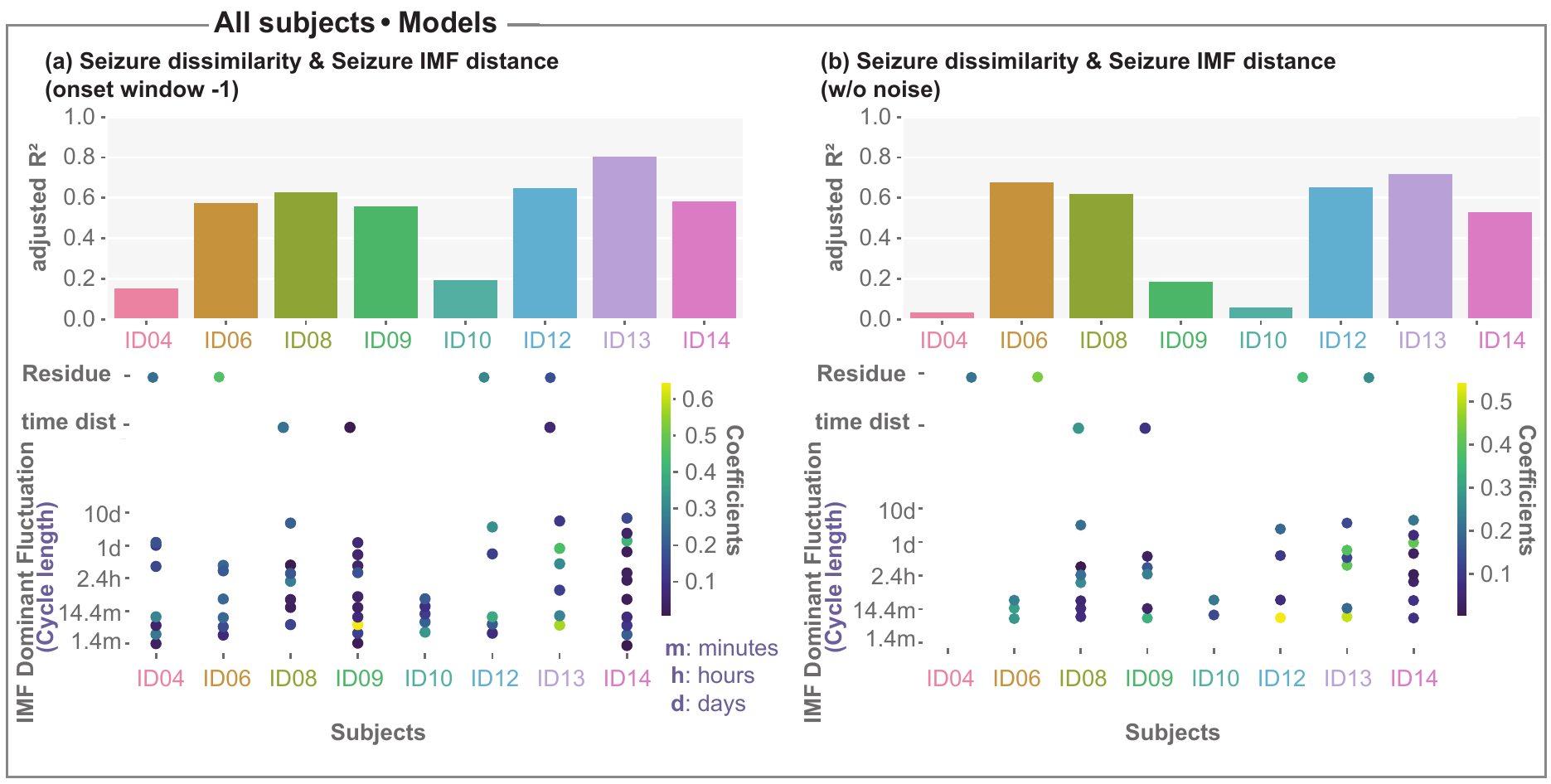}
       \caption{(a)\&(b) Summary across subjects based on OLS models with explanatory variables obtained by the constrained LASSO using similar representation as in Fig.\ref{fig:lasso}d for the models described in Section \ref{subsec:raw_analysis_onset-1} (Left plot: (a)) and Section \ref{subsec:Raw_analysis-noise} (Right plot: (b)). Top: Bar chart of the adjusted $R^2$. Bottom: Dot plot indicating the OLS coefficient estimates for the residue or time distance (when this variable remained in the model) together with OLS coefficient estimates at the corresponding value of IMF peak frequency for each subject. For visualisation, we converted the peak frequency to cycle length.}
    \label{fig:R_squared_all_models}
\end{figure}

We further performed the regression analysis in Section~\ref{subsec:Linear analysis} excluding the first three IMFs, which could represent noise (\ref{subsec:noisyIMF}). As can be seen in Fig.~\ref{fig:R_squared_all_models}, the adjusted $R^2$ values for the majority of subjects were comparable for the two models (albeit generally slightly lower). Only for subjects ID09, the adjusted $R^2$ was dramatically lower for the model without the first three IMFs (see Table~\ref{tab:R^2_Raw_Analysis} and Fig.~\ref{fig:R_squared_all_models}b). Note also our Discussion on the role of the faster IMFs.

We also observed that the IMFs and corresponding coefficients were substantially different in ID04, ID09, and ID10 between the two models. This is not surprising given that both ID04 and ID10 had a low adjusted $R^2$ in the first place, and ID09 had a low adjusted $R^2$ in the model without the first IMFs. In summary, we conclude that the first three IMFs do not contribute substantially to explaining seizure dissimilarities in most subjects. However, in some subjects, faster IMFs may play a strong role in explaining seizure dissimilarities. 


\begin{table}[h]
\centering
\begin{tabular}{cccc}
\textbf{subjects}                                                              & \multicolumn{3}{c}{\textbf{\begin{tabular}[c]{@{}c@{}}Seizure dissimilarity \& IMF Seizure distance\\ Adjusted $\boldsymbol{R^2}$\end{tabular}}}                                                         \\ \hline
\multicolumn{1}{|l|}{ \textbf{}} & \multicolumn{1}{c|}{\textbf{onset window}} & 
\multicolumn{1}{c|}{\textbf{onset window -1}} & \multicolumn{1}{c|}{\textbf{\begin{tabular}[c]{@{}c@{}}onset window\\ w/o noise\end{tabular}}} \\ \hline
\multicolumn{1}{|c|}{\textbf{ID04}} & \multicolumn{1}{c|}{0.3343}                                      & \multicolumn{1}{c|}{0.1545}                                              & \multicolumn{1}{c|}{0.0347}                                                 \\ \hline
\multicolumn{1}{|c|}{\textbf{ID06}}  
              & \multicolumn{1}{c|}{0.6742}                                & \multicolumn{1}{c|}{0.5755}                              
              & \multicolumn{1}{c|}{0.6787}                                  \\ \hline
 
\multicolumn{1}{|c|}{\textbf{ID08}}                    & \multicolumn{1}{c|}{0.6207}                                      & \multicolumn{1}{c|}{0.6279}                                         & \multicolumn{1}{c|}{0.6201}                                                                                          \\ \hline

\multicolumn{1}{|c|}{\textbf{ID09}}                    & \multicolumn{1}{c|}{0.5927}                                      & \multicolumn{1}{c|}{0.5601}                                         & \multicolumn{1}{c|}{0.1865}                                                                                          \\ \hline
 
\multicolumn{1}{|c|}{\textbf{ID10}}                    & \multicolumn{1}{c|}{0.1419}                                      & \multicolumn{1}{c|}{0.1949}                                         & \multicolumn{1}{c|}{0.0608}                                                                                           \\ \hline
 
\multicolumn{1}{|c|}{\textbf{ID12}}                    & \multicolumn{1}{c|}{0.7299}                                      & \multicolumn{1}{c|}{0.6489}                                         & \multicolumn{1}{c|}{0.6529}                             \\ \hline
\multicolumn{1}{|c|}{\textbf{ID13}}                    & \multicolumn{1}{c|}{0.8243}                                      & \multicolumn{1}{c|}{0.8060}                                         & \multicolumn{1}{c|}{0.7197}                             \\ \hline     

\multicolumn{1}{|c|}{\textbf{ID14}}                    & \multicolumn{1}{c|}{0.5973}                                      & \multicolumn{1}{c|}{0.5834}                                         & \multicolumn{1}{c|}{0.5306}                                                                                          \\ \hline
\end{tabular}
\caption{\label{tab:R^2_Raw_Analysis}Adjusted $R^2$ values for the models described in Sections~\ref{subsec:Linear analysis},~\ref{subsec:raw_analysis_onset-1} \& \ref{subsec:Raw_analysis-noise} for each subject with at least six recorded seizures.}
\end{table}

\end{document}